\documentclass[onecolumn]{aa} % for a paper on 1 column  
\usepackage{graphicx}
\usepackage{txfonts}
\usepackage{natbib}
\usepackage{amssymb}

\newcommand{\hess}{H.E.S.S.}
\newcommand{\crab}{Crab\,Nebula}
\newcommand{\cena}{Centaurus\,A}
\newcommand{\gc}{HESS\,J1745--290}
\newcommand{\hessjj}{HESS\,J1507--622}
\newcommand{\vjr}{Vela\,Junior}
\newcommand{\vx}{Vela\,X}

\begin{document}
   \title{A new method of reconstructing VHE $\gamma$-ray spectra: \\the Template Background Spectrum}
   \author{\thanks{milton.virgilio.fernandes@physik.uni-hamburg.de}\, Milton\,Virg{\'{\i}}lio \,Fernandes$^1$,
          Dieter\,Horns$^1$,
          Karl\,Kosack$^2$,
          Martin\,Raue$^1$,
          \and Gavin\,Rowell$^3$}
   \institute{$^1$Institut f\"ur Experimentalphysik, Universit\"at Hamburg, Luruper Chaussee 149, D 22761 Hamburg, Germany\\
              $^2$DSM/Irfu, CEA Saclay, F-91191 Gif-Sur-Yvette Cedex, France\\
              $^3$School of Chemistry and Physics, University of Adelaide, Adelaide 5005, Australia              
	                     }
   \date{Received -/-; accepted -/-}

% \abstract{}{}{}{}{} 
% 5 {} token are mandatory
 
  \abstract
   {Very-high-energy (VHE, $E>0.1$\,TeV) $\gamma$-ray emission regions with angular extents comparable to the field-of-view of current imaging air-Cherenkov telescopes (IACT) require additional observations of source-free regions to estimate the background contribution to the energy spectrum. This reduces the effective observation time and deteriorates the sensitivity.}
   {A new method of reconstructing spectra from IACT data without the need of additional observations of source-free regions is developed. Its application is not restricted to any specific IACT or data format.}
   {On the basis of the \emph{template} background method, which defines the background in air-shower parameter space, a new spectral reconstruction method from IACT data is developed and studied, the Template Background Spectrum (TBS); TBS is tested on published \hess\, data and \hess\, results.}
   {Good agreement is found between VHE $\gamma$-ray spectra reported by the H.E.S.S. collaboration and those re-analysed with TBS. This includes analyses of point-like sources, sources in crowded regions, and of very extended sources down to sources with fluxes of a few percent of the Crab Nebula flux and excess-to-background ratios around 0.1. However, the TBS background normalisation introduces new statistical and systematic errors which are accounted for, but may constitute a limiting case for very faint extended sources.}
   {The TBS method enables the spectral reconstruction of data when other methods are hampered or even fail. It does not need dedicated observations of VHE $\gamma$-ray-free regions (e.g. as the \textit{On}/\textit{Off} background does) and circumvents known geometrical limitations to which other methods (e.g. the reflected-region background) for reconstructing spectral information of VHE $\gamma$-ray emission regions are prone to; TBS would be, in specific cases, the only feasible way to reconstruct energy spectra.}

\authorrunning{Fernandes et al.}
\titlerunning{Template Background Spectrum}

\maketitle
%
%________________________________________________________________

\section{Introduction}\label{section:intro}
Since the first detection of a very-high-energy (VHE, $E>0.1$\,TeV) $\gamma$-ray source \citep[the \crab;][]{Whipple} more than two decades ago, the VHE $\gamma$-ray regime has established itself as an important astrophysical observation window. 
Data from ground-based imaging air-Cherenkov telescope systems (IACTs), e.g. \hess\,, MAGIC, and VERITAS \citep[recently reviewed in ][]{Hillas2013} are essential to the understanding of acceleration and emission processes present at various astrophysical objects. 
However, VHE $\gamma$-rays are not directly observable with IACTs because they interact in the upper layers of the atmosphere to produce relativistic particle cascades (extensive air showers) emitting Cherenkov light that can then be recorded with IACTs. 
These recorded images are used to reconstruct the relevant properties of the incident primary particle. 
However, the observed flux contains not only $\gamma$-ray-induced air showers, but also hadron-induced ones which dominate the former at these energies. 
Hence, VHE $\gamma$-ray events are not directly accessible before the hadronic background is accurately determined.

A simple but powerful way to distinguish between the two regimes has been proven to be the characterisation of these shower images by the first moments of the intensity distribution because the hadron-induced images are irregularly and broader shaped \citep{HillasTech}. 
Additionally, the stereoscopic recording of an air-shower event by two or more telescopes further reduces the hadronic contamination \citep{HEGRA}. 
In the last five years, the performance in the IACT data analysis has been improved by more sophisticated methods exploiting all information of the accessible air-shower parameters.
These sophisticated methods cover a wide range of approaches and techniques ranging from machine-learning algorithms, neural networks, and likelihood approaches to semi-analytical combinations of these approaches \citep{Model,TMVA,Model3D,Dubois2009,Fiasson2010,MAGIC_randomforest}. 
However, despite these methods the hadronically-induced air-shower events (i.e. the background) still constitute the major fraction of the data. 
Therefore, further analysis techniques are applied.

In contrast to most extra-Galactic VHE $\gamma$-ray emitters, Galactic sources tend to be more extended than the point-spread function (PSF) of the IACT. 
Some of them exhibit angular sizes approaching the field-of-view (FoV). 
As observations in the VHE $\gamma$-ray regime are highly background-dominated, additional observation time under similar observation conditions of VHE $\gamma$-ray emission-free regions is required to estimate the background. 
This approach, however, is not favoured as it is at the expense of available and limited dark time. A simultaneous background determination to reconstruct spectral information of extended sources could also be useful for the next-generation IACTs, e.g. the Cherenkov Telescope Array \citep[CTA,][]{CTASurvey}.

State-of-the-art background estimation methods \citep[e.g. reviewed in][]{BgTechs} cannot handle both large source extents and limited FoVs to determine source spectra. 
In this work, a new method for reconstructing spectral information from (extended) VHE $\gamma$-ray sources without the expense of additional observation time is presented. 
This method, the Template Background Spectrum (TBS), is not restricted to specific IACT data or setup. 
For testing purposes only, the high-quality \hess\, data were used. 
Therefore, parameter ranges and cut values (e.g. for the zenith angle or the background rejection) stated in this work apply only to these data. However, an adjustment of these experiment-related values should suffice for the analysis of data obtained with other instruments.

In Sect.\,\ref{section:backgrounds}, a short overview of the relevant background techniques is provided. Then the new method for imaging spectroscopy is introduced (Sect.\,\ref{section:method}) and its results are compared with published \hess\, spectra (Sect.\,\ref{section:discussion}) before we conclude in Sect.\,\ref{section:summary}.

\section{Background estimation methods}\label{section:backgrounds}
The stereoscopic approach pioneered with the HEGRA IACT system \citep{HEGRA} is an efficient way to suppress the hadronic background, but even then ground-based observations in the VHE $\gamma$-ray regime are still highly background-dominated and therefore further procedures have to be taken to extract a signal. 
Before we briefly elucidate the common background-estimation methods, some discussion is given particularly on the camera acceptance and on the so-called $\gamma$/hadron separation.

\paragraph{Camera acceptance:} The systemic response for the detection of $\gamma$-ray events changes over the FoV. 
In general, events are detected towards the camera centre with a higher efficiency/probability than those at or close to the camera edge thus making it unfavourable for observing sources at large offsets (i.e. distances  from the camera centre). 
This effect is similar to vignetting.

This systemic response, the camera acceptance, drops off radially with increasing angular distance $\theta$ with respect to the optical axis\footnote{In most cases, this is a good approximation.} and it is different for VHE $\gamma$-rays and the (hadronic) background. 
It also depends on energy and zenith angle of the incoming particle, and at low energies, also on the alignment with respect to the Earth's magnetic field. 
One assumes the camera acceptance to be symmetric with respect to the azimuth angle.

\paragraph{Gamma/Hadron separation:} As addressed earlier, hadronic air-shower events constitute a large fraction of the data. 
Hence, criteria are needed to distinguish between real $\gamma$-ray events and hadron-like ones. 
One discriminates between the two regimes on the basis of at least one parameter with a high separation power. 
This parameter may be just based on some (inferred) observable or expectation or the result of machine-learning/multi-variate algorithms that assign a $\gamma$/hadron-likeness to each event. 
The separation parameter is essential for the template background (introduced later in this section) and thereby also for the Template Background Spectrum. 
However, their application to data does not depend on the separating parameter itself. 
In the remainder of this work, the Hillas air-shower parameter \textit{width} is discussed and used to separate $\gamma$-ray events from hadron-like ones.

\citet{HillasTech} demonstrated that shower images can be described by their first moments. 
Among these, the width has proven to be the parameter with the strongest separation power between VHE $\gamma$-ray and hadronic air showers. 
The transverse momentum of composite particles lead to a wider image in the camera.

From instrument-specific Monte Carlo (MC) simulations of $\gamma$-rays, an expected value for the width of a VHE $\gamma$-ray-induced air shower $w_\mathrm{MC}$ and its spread $\sigma_\mathrm{MC}$ for different observational conditions (image amplitude, zenith angle, impact parameter) are known. 
After characterisation of the air-shower image, its telescope-wise reduced width $w_i$ scaled with respect to the MC expectation is
\begin{equation}\label{eqn:scw}
 w_{\mathrm{rsc},i} = \frac{w_i-w_\mathrm{MC}}{\sigma_\mathrm{MC}}\,\mathrm{.}
\end{equation}
The mean reduced scaled width (MRSW) of all participating $N$ telescopes is then 
\begin{equation}\label{eqn:mrsw}
 \mathrm{MRSW} = \frac{\sum_i^N w_{\mathrm{rsc},i}}{N}\,\mathrm{.}
\end{equation}
The MRSW parameter follows a $N(0,1)$ distribution and therefore should peak around 0 for VHE $\gamma$-ray photons.
Events within a certain range around 0 will be classified as \textit{gamma}-like. 
Hadronic air-shower events leave a more irregular and wider imprint in the camera leading to higher MRSW values. 

For most background-estimation methods, only the VHE gamma-like events are considered and all other events are discarded. 
However, as known from MC simulations of protons, this gamma-like sample does still include a majority of hadronic events. 
Hence, the exact $\gamma$-like range in MRSW is a compromise between a rather clean sample consisting of a comparatively large VHE $\gamma$-ray fraction and large overall statistics with a higher hadronic contamination. 

In principle, this contamination can be determined through MC simulations as done in \citet{ElectronPaper} where the contribution of cosmic-ray electrons to the $\gamma$-like sample were estimated through means of further electron MC simulations. 
However, this approach would invoke extensive simulations of cosmic-ray hadrons. Besides, the uncertainty of the hadronic interaction models would translate into a non-negligible systematic error. 
Therefore, MC simulations to determine the hadronic background in data are generally not used.

\paragraph{Estimating the excess events:} To estimate the hadronic contamination within the gamma-like distribution in the source region of interest (often called ON region), one defines a background control region (often called OFF region) to estimate the background. 
The observed VHE $\gamma$-ray excess $N_\mathrm{excess}$ would then be 
\begin{equation}\label{eqn:std_excess}
 N_\mathrm{excess} = N_\mathrm{ON}-\alpha N_\mathrm{OFF}\,\mathrm{,}
\end{equation}
where $N_\mathrm{ON}$ and $N_\mathrm{OFF}$ are the number of events passing the same cuts in the respective regions with an overall normalisation $\alpha$ that accounts for any (observational) difference between ON and OFF, e.g. solid angle or exposure. 
For spectral studies, $N_\mathrm{ON}$ and $N_\mathrm{OFF}$ are subdivided into energy intervals. 
The aim of any background-determination method is the estimation of $\alpha$ and $N_\mathrm{OFF}$. The relative statistical uncertainty can be reduced by increasing $N_\mathrm{OFF}$ (decreasing $\alpha$). 
In general, VHE $\gamma$-ray observations are usually conducted as a sequence of short time spans (often called \textit{runs}\footnote{With \hess, these runs are usually around 30\,mins long.}) and the excess events are normally also calculated on a run-by-run basis.

In the remainder of this section the two commonly used methods to estimate the background in energy spectra, namely the \textit{On}/\textit{Off} background and the reflected-region background, are introduced.
Furthermore, the template background model is presented which is hitherto only in use for skymap generation and source detection, but will serve as the background estimator for TBS.
For more details on these methods than presented in the following, we refer the interested reader to \citet{Whipple,HEGRA,TemplateBg,Crab_paper}; and \citet{BgTechs}.

\subsection{\textit{On}/\textit{Off} background} 
Observations of VHE $\gamma$-ray source-free regions are conducted to estimate the OFF sample. 
These non-simultaneous observations are performed under identical conditions (e.g. same altitude and azimuth angle) and should not be too separated in time because of possible changing weather conditions (during observation night and seasonal differences) and the degradation of the optical elements of any telescope system (on timescales of months to years). 
Ideally, one alternates between \emph{On} and \emph{Off} observations to assure data under the same conditions. 
If the optical degradation is accounted for, a data base can be used to match \emph{On} and \emph{Off} observations separated by several months. 
This, however, introduces new systematic effects if the match is not perfect. 
The \textit{On}/\textit{Off} background is normally in use for single-telescope systems, e.g. Whipple \citep{Whipple}, for morphological and spectral studies and when other background estimates are not applicable. 
However, in order to achieve $\alpha\approx1$, double the amount of dark time has to be spent on dedicated \textit{On}/\textit{Off} observations. This background method can be used for both morphological and spectral studies.

\subsection{\textnormal{Wobble} observations and reflected-region background}\label{subsection:wobble} 
The reflected-region background is a consequence of the so-called \textit{wobble(d)} observations.
The camera centre is not pointed at the source of interest, but shifted (\emph{wobbled}) by an \textit{a\,priori} chosen angle $\omega$ in right ascension or declination. 
It was first conducted by the HEGRA experiment \citep{HEGRA}. 
During a campaign on a specific source consisting of observations of equal time spans, the wobble positions are altered such that the exposure in the FoV is uniform around the ON region. 
Now, OFF regions share the same properties as the ON region as they are placed at the same radial distance from the camera centre (\textit{reflected} with respect to the camera centre and ON region; avoiding VHE $\gamma$-ray contamination) and therefore exhibit the same camera acceptance. 
For equally-sized ON and OFF, the normalisation is simply $\alpha = 1/n_\mathrm{OFF}$ where $n_\mathrm{OFF}$ is the number of OFF regions.\footnote{Alternatively, one can also use a ring segment placed at the same radial distance; here, $\alpha$ is then the ratio of the solid angles.} 
This simple approach keeps systematic effects as low as possible making it a reliable technique for spectral studies. 
In case of VHE $\gamma$-ray emission (diffuse or from sources) in the FoV $n_\mathrm{OFF}$ is reduced. 
However, if the radius of the ON region $\theta_\mathrm{ON}$ is larger than $\omega$, no OFF region can be placed. 

\subsection{Template background}\label{section:template} 
The template background \citep{TemplateBg} was designed for source detection and morphological studies. 
For this, a sample of background events is used which is normally discarded from the data in the methods described above and now serves as a template to model the background. 
For a clearer discrimination from the previous methods, the source region of interest is now called the \textit{signal} region and the remainder, excluding VHE $\gamma$-ray emission regions, is the FoV sample. 
Both samples are divided with respect to the $\gamma$/hadron-separation parameter into a gamma-like sample (as in the other methods) and a hadron-like sample. 
Both gamma-like samples (events from the signal region and from the FoV passing the $\gamma$-like cut) and both hadron-like samples (events from the signal region and from the FoV passing the hadron-like cut) will experience a different camera acceptance. 
This difference is corrected by a higher-order polynomial fit $P$ of the gamma-like (g) and the hadron-like (h) sample as a function of the angular distance from the camera centre $\theta$. 
Additionally, the zenith-angle dependence $f(z)$ over the FoV has to be accounted for:
\begin{equation}\label{eqn:std_alpha_theta}
 \alpha(\theta,z) = \frac{P_\mathrm{g}^\mathrm{FoV}(\theta)}{P_\mathrm{h}^\mathrm{FoV}(\theta)}\,f(z)\,.
\end{equation}

With this template-background normalisation, the excess $N_\mathrm{excess}$ at a given position in the FoV is
\begin{equation}\label{eqn:std_excess_alpha_theta}
 N_\mathrm{excess}(\theta) = N_\mathrm{g}(\theta)-\alpha(\theta,z)N_\mathrm{h}(\theta)\,, 
\end{equation}
where $N_\mathrm{g}$ and $N_\mathrm{h}$ are the gamma-like and hadron-like events, now also including source region and other VHE $\gamma$-ray emission regions.
The advantage of this method is that local effects, e.g. stars in the FoV or defective pixels in the camera, are accounted for as they appear in both regimes at the same location and are virtually cancelled out. 
Moreover, this template background can be used for large source radii and is much less restricted by geometrical limitations in the FoV (Sect.\,\ref{subsection:wobble}). 
However, for large source regions, the camera acceptance has to be extrapolated since large parts of the FoV have to be excluded. 
A pile-up of reconstructed events towards the edge of the FoV is known to occur and is due to truncated camera images at the FoV edge \citep{TemplateBg}. This can be mitigated by rejecting events close to the edge of the camera. 
Therefore, the application of the template background is limited to the inner $\sim90\,\%$ of the full FoV. 
Already for $\sim10$\,hrs of data, the relative statistical errors on $\alpha$ are below 1\,\% and therefore negligible.

Compared to the other background estimation methods, the template normalisation $\alpha$ is not calculated by simple geometrical means but requires a good knowledge of both acceptances.
Yet, correcting the data with this $\alpha$ introduces another potential source of a systematic error related to the accuracy of its computation in Eq. \ref{eqn:std_alpha_theta} and of the azimuthal symmetry. 
Additionally, after excluding the signal region and other known VHE $\gamma$-ray emission regions, there have to be sufficient data in the FoV to calculate $\alpha(\theta,z)$. 

\subsection{Summary}
In general, the reflected-region background in combination with \textit{wobble} observations is used to extract spectral information.
If this is not possible the \textit{On}/\textit{Off} background is used. 
If none of them is applicable, spectral information from the source is partly lost by discarding data without a good background estimate. 

In the light of the more sensitive telescope array CTA, it is expected that the FoVs along the Galactic Plane are much more crowded with probably several new $\gamma$-ray sources and therefore providing fewer suitable data in the FoV to apply the reflected-region background.
Hence, an alternative background estimate for energy spectra is motivated.

As it is usually always possible to create skymaps with the template background model, this background estimate is used to reconstruct spectral information and to overcome the limitations of the reflected region method.
\section{Template Background Spectrum (TBS)}\label{section:method}
In order to be able to estimate the background in energy spectra based on the template background model, the following aspects have to be considered.

First of all, the template normalisation $\alpha$ (Eq. \ref{eqn:std_alpha_theta}; henceforth this definition is referred to as $\alpha_\mathrm{std}$) is averaged over all energies $E$, which is sufficient for skymap generation and source detection but not appropriate for reconstructing (energy) spectra. 
Hence, we propose and introduce the energy dependence into $\alpha(z,\theta)$ : 
\begin{equation}\label{eqn:tbs_alpha}
 \alpha(E,z,\theta) =  \frac{N_\mathrm{g}^\mathrm{FoV}(E,z,\theta)}{N_\mathrm{h}^\mathrm{FoV}(E,z,\theta)}\,;
\end{equation}
where $N_\mathrm{g}^\mathrm{FoV}$ and $N_\mathrm{h}^\mathrm{FoV}$ are the event distributions of both regimes in the FoV. 
From here on and if not stated otherwise, $\alpha$ always implies the full dependency $\alpha(E,z,\theta)$.

Secondly and already mentioned earlier, the camera acceptance behaves differently for gamma-like and hadron-like events. 
Moreover, these individual acceptances are subject to energy and zenith-angle dependent changes \citep[discussed in detail in][]{BgTechs}.
It has been suggested that these relative differences lead to systematic effects hampering a spectral reconstruction based on the template background model \citep[e.g.][]{BgTechs}.
However, the issue of the different relative acceptances including energy and zenith-angle dependencies is intrinsically accounted for because the template correction $\alpha$ is defined as the ratio of gamma-like and hadron-like events (Eq. \ref{eqn:tbs_alpha}) and thus already accounts, on average, for the relative differences within $(E,z,\theta)$.

The third aspect is the lack of sufficient statistics to calculate $\alpha$ on a runwise basis. 
This occurs when a major fraction of the data have to be excluded, either because of many sources in the FoV or because of the large signal region itself.
To circumvent this issue, a so-called \textit{lookup} of the template correction is created (described in Sect.\,\ref{subsection:lookup}).
 
In the following, \hess\, data are used for the general study of the template correction.

\subsection{Lookup creation}\label{subsection:lookup}
After excluding all VHE $\gamma$-ray emission regions, the remaining FoV data are split into intervals of energy, zenith angle, and camera offset, but will often not be sufficient to determine $\alpha$ in Equation \ref{eqn:tbs_alpha} on a run-by-run analysis.
To circumvent this problem, the events from the FoV of all observations to be spectrally analysed (i.e. no \textit{Off} data are used) are used to create a \textit{lookup table} binned in energy, zenith angle, and camera offset.
This lookup table consists of the accumulated distributions $N_\mathrm{g}^\mathrm{FoV}(E,z,\theta)$, $N_\mathrm{h}^\mathrm{FoV}(E,z,\theta)$, and the inferred $\alpha(E,z,\theta)$ calculated through Equation \ref{eqn:tbs_alpha}.
The binning has to be chosen to assure sufficient statistics to calculate $\alpha$.
The signal-region data are binned identically. 
The exact choice of range and bin width in energy, zenith angle, and offset are to a certain extent arbitrary and IACT-specific. 

This lookup requires a good knowledge of the parameters, especially the energy which involves two aspects in the IACT data analysis.
On the one hand, the energy resolution deteriorates at lower energies. 
This can lead to an over- or underestimation of $N_\mathrm{g}^\mathrm{FoV}$ and $N_\mathrm{h}^\mathrm{FoV}$ in the lookup and thus to an incorrect estimate of $\alpha$.
As any event-wise information other than energy, zenith angle, and offset is lost when the lookup is filled with the data from different runs, events with a poor energy resolution have to be rejected \textit{a priori}.\footnote{When determining the best-fit spectrum (presented later), folding or unfolding techniques only account for the expected $\gamma$-like events per observation, but not for the background normalisation which is calculated after accumulating data from different runs.} 
For this, an energy-threshold cut is applied.
The energy threshold is determined through the relative energy bias $E_\mathrm{bias}$. 
The relative bias is the normalised deviation between the energies of reconstructed and simulated events, $E_\mathrm{reco}$ and $E_\mathrm{sim}$, respectively:
\begin{equation}\label{eqn:eth}
 E_\mathrm{bias}(z_\mathrm{sim},\omega_\mathrm{sim}) =  \left|\frac{E_\mathrm{reco}-E_\mathrm{sim}}{E_\mathrm{sim}}\right|(z_\mathrm{sim},\omega_\mathrm{sim})\,\mathrm{TeV}\,.
\end{equation}
It is determined in MC simulations for different run zenith angles $z_\mathrm{sim}$ and pointing offsets $\omega_\mathrm{sim}$. 

On the other hand, the performance of any IACT degrades with time inevitably leading to a false energy reconstruction.
If not accounted for, events with a false reconstructed energy will be not filled into the correct $(E,z,\theta)$ bin of the lookup.
This problem becomes more severe if the individual runs used in the analysis are separated on larger timescales of about a year or more.
Hence, a good estimate of this loss in efficiency over time is required to avoid an energy bias affecting the lookup.
This is achieved through the analysis of recorded muon rings in the IACT cameras \citep{MAGIC_muon,Bolz2004,VERITAS_muon}, and the correction of the reconstructed energies can be done sufficiently well by a shift in energy proportional to the loss in efficiency.

Bright stars in the FoV are not a major concern in the standard runwise application of the template background \citep{TemplateBg}.
However, it was found that this background method slightly overcorrects the background and creates a slight positive significance at the star position \citep{BgTechs}.
In a conservative approach and also because of the lookup creation, stars with an apparent magnitude of 5 or brighter are excluded in TBS.
However, for TBS, any effect would likely be smeared out. 
Given a normal wobble-observation pattern around the source of interest, stars are in general never to be found at the very same position with respect to the camera centre and therefore located at a different offset $\theta$. 
Hence, any effect would be difficult to be traced back and to be quantified in $(E,z,\theta)$ space.

\subsection{$(E,z,\theta$) dependence of $\alpha$}
\begin{figure*}[t]
  \resizebox{\hsize}{!}{
    \includegraphics{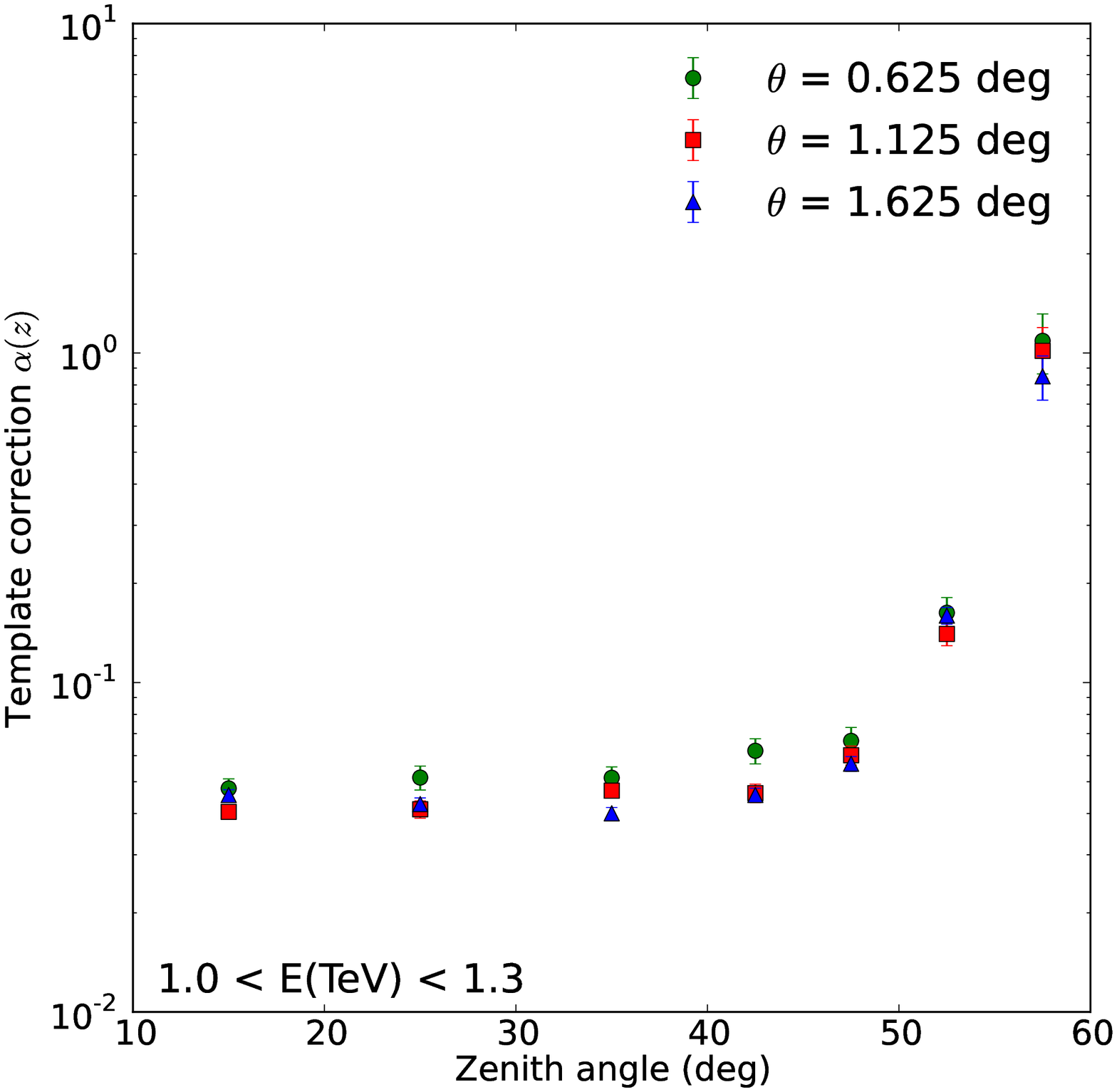}
    \includegraphics{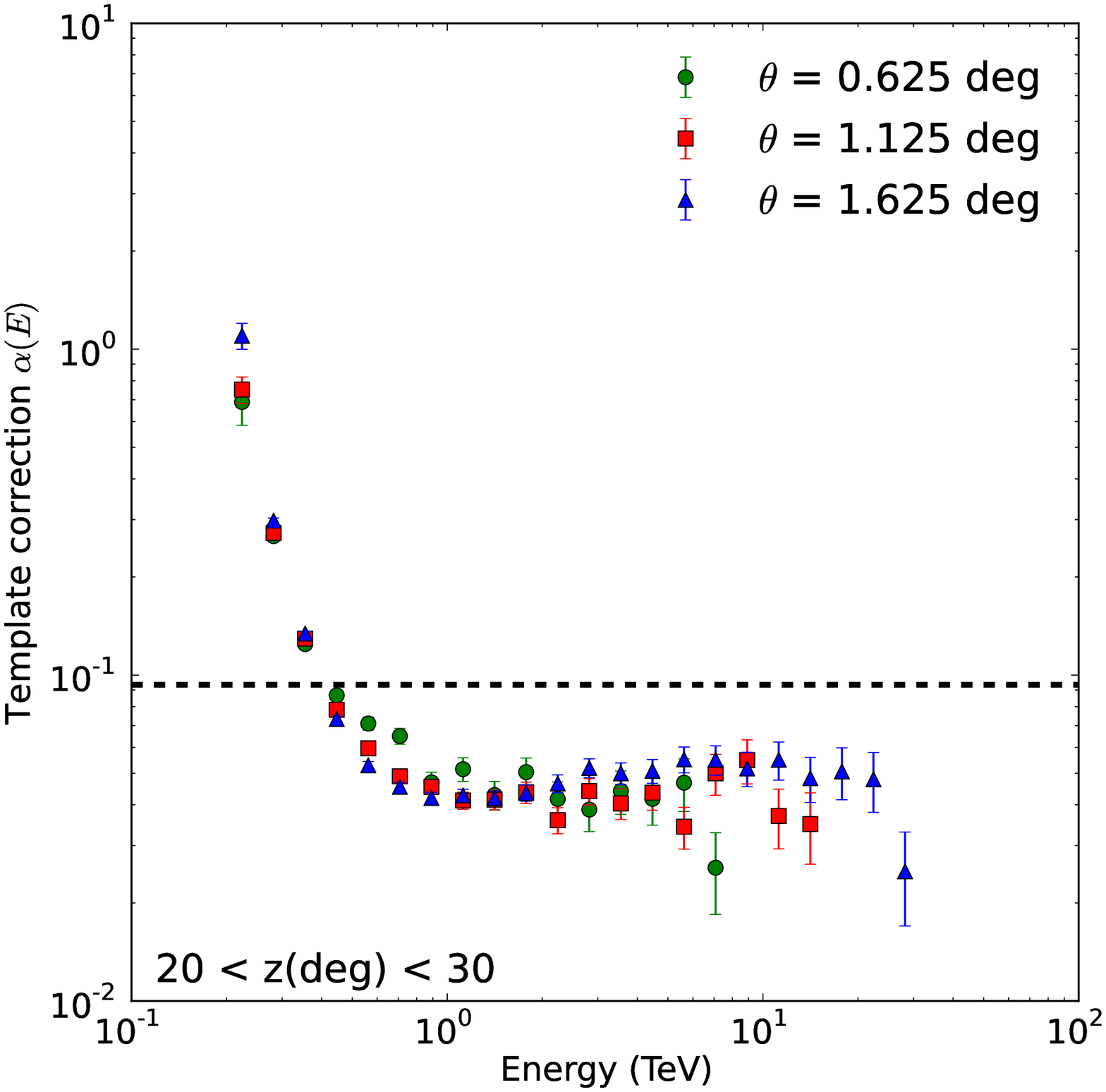}}
  \caption{TBS correction $\alpha(E,z,\theta)$ calculated from \hess\, data on \gc\, according to Eq. \ref{eqn:tbs_alpha} for different offsets $\theta=0.625^\circ$ (green circles), $1.125^\circ$ (red squares), $1.625^\circ$ (blue triangles). \emph{Left:} Zenith-angle dependence of $\alpha$ calculated for the energy interval $1<{E}/\mathrm{TeV}\leq1.3$. \emph{Right:} Energy dependence of $\alpha$ calculated for the zenith-angle interval $20^\circ<{z}\leq30^\circ$. The black-dashed line indicates the energy-averaged template correction. See text for further information.}
  \label{fig:tbs_EZ}
\end{figure*}
In Fig.\,\ref{fig:tbs_EZ}, the dependencies of the TBS correction are illustrated. For this, \hess\, data on the Galactic Centre \citep[\gc,][]{GC_paper} were used (the analysis of this source will be discussed later).

In the left-hand plot, the trend of $\alpha$ for increasing zenith angles for the energy range $1<{E}/\mathrm{TeV}\leq1.3$ and three values of the camera offset $\theta$ is shown; $\alpha$ is comparatively constant up to zenith angles of about $40^\circ$ but afterwards quickly approaches unity at high zenith angles. 
In the right-hand plot, $\alpha$ is plotted against energy for the zenith-angle interval $20^\circ<{z}\leq30^\circ$. 
The correction is around unity for the lowest energies and then drops to saturate at energies beyond 1\,TeV. 
Here, events at lower offset do not reach beyond $\sim10\,$TeV because of the limited FoV (discussed in the following).
Compared to the energy-independent $\alpha_\mathrm{std}$ (dashed line), the template correction is significantly underestimated at lower energies and overestimated at higher energies (right-hand side of Fig.\,\ref{fig:tbs_EZ}).

In Fig.\,\ref{fig:tbs_EZ} the behaviour of $\alpha$ is presented for a fixed energy range ($1<{E}/\mathrm{TeV}\leq1.3$) and a fixed zenith-angle range ($20^\circ<{z}\leq30^\circ$). 
At lower zenith angles and higher energies the camera images are brighter, thus leaving more light to reconstruct the shower parameters and therefore to separate gamma-like and hadron-like events. 
As a result, $\alpha\ll1$. 

At the same energy, hadron-induced air showers are not as bright as the gamma-like ones.
Hence, at lower energies and at higher zenith angles where the images are fainter, the hadron-like events are affected more strongly than the $\gamma$-like events.
Comparatively fewer hadrons are reconstructed leading to higher value of $\alpha$.

At higher energies, air showers tend to be reconstructed towards higher camera offsets. 
This is because the respective acceptance close to the camera centre is comparatively low. 
The individual shower images exhibit a larger distance (image centre-of-gravity to camera centre).
The higher the energy the larger this distance, until the images are truncated or ultimately lie outside the camera.

\subsection{TBS excess events}
In the following, the calculation of the TBS excess events and the respective errors are presented. 
In contrast to the reflected-region background or the \textit{On}/\textit{Off} background (Sect.\,\ref{section:backgrounds}), an event-based correction has to be determined in the three-dimensional parameter space $(E,z,\theta)$ through interpolation or extrapolation of the lookup table. 
A detailed description, also involving a discussion of different systematic effects, is given in Appendix \ref{appendix:tbs}. 

Every hadron event $i$ from the signal region with its specific properties $(E_i,z_i,\theta_i)$ is weighted with a value $\beta_i({\Delta}E,{\Delta}z,\theta_i)$, where $\Delta{E}$ and $\Delta{z}$ are the energy and zenith-angle bin, respectively, and $\beta_i$ is calculated from the three-dimensional lookup. 
If $m$ is the total number of hadron events within an energy bin $\Delta{E}$, the respective TBS excess is
\begin{equation}
 N_\mathrm{excess}(\Delta{E}) = N_\mathrm{g}^\mathrm{s}(\Delta{E}) - \sum_i^m\beta_i(\Delta{E},\Delta{z}_i,\theta_i)\,,
\end{equation}
and with Equation \ref{eqn:tbs_beta_eff},
\begin{equation}\label{eqn:gamma_tbs}
 N_\mathrm{excess}(\Delta{E}) = N_\mathrm{g}^\mathrm{s}(\Delta{E}) - \beta_\mathrm{eff}(\Delta{E})N_\mathrm{h}^\mathrm{s}(\Delta{E})\,,
\end{equation}
where $\beta_\mathrm{eff}$ is the effective TBS correction. 
The corresponding error on the excess counts is then
\begin{equation}\label{eqn:gamma_tbs_error}
 \sigma(N_\mathrm{excess}) = \sqrt{N_\mathrm{g}^\mathrm{s} + \beta_\mathrm{eff}^2N_\mathrm{h}^\mathrm{s} + (\sigma(\beta_\mathrm{eff})N_\mathrm{h}^\mathrm{s})^2}\,.
\end{equation}
The additional term $(\sigma(\beta_\mathrm{eff})N_\mathrm{h}^\mathrm{s})^2$ accounts for the statistical errors of the TBS correction and includes the uncertainties on interpolation and extrapolation (discussed in Appendix \ref{appendix:tbs}). 
Its contribution to the overall error $\sigma(N_\mathrm{excess})$ depends on the data of the analysed source itself. 
In general, the introduction of a third term to the error calculation will lead to higher statistical errors on the excess counts compared to those of reflected-region background or \textit{On}/\textit{Off} background. 
This additional contribution to the error will be discussed on the basis of the analysed \hess\, data sets in Sect.\,\ref{section:discussion}.

\subsection{Parameter estimation}
The spectral-parameter estimation of an assumed spectral behaviour $dN/dE$ is done by a forward-folding technique based on a $\chi^2$ minimisation similar to \citet{XSpec}. 

The excess counts $N_\mathrm{pred}$ for an observation are predicted by taking into account the full instrument response of the IACT. 
The instrument response is usually well known and determined in MC simulations. 
This response is, in general, described by the effective area and the energy resolution.
For TBS, the \hess\, MC data were reprocessed to produce the effective areas and energy-resolution matrices matching the TBS cuts and parameter ranges.

The binned MC data were produced for different zenith angles, wobble offsets, and a wide energy range. 
As the optical parts of the telescope are subject to degradation over time, the loss in efficiency is estimated through single muon rings \citep{Bolz2004}. 
This is done by a shift in energy proportional to the estimated loss in efficiency for the effective area and the energy threshold.
For the energy-resolution matrix this shift is done along the diagonal of the matrix (since the energy-resolution matrix can be considered a diagonal matrix in a simplified scheme).

Information on the \hess\, MC data can be found in \citet{Crab_paper}.

\section{Application to \hess\, data}\label{section:discussion}
In this section \hess\ data are used. 
On their basis, parameter ranges and values are set to test the new method for estimating background for VHE $\gamma$-ray spectra. 
However, for application to other IACT data, it should be sufficient to just replace the values related to the \hess\, instrument. 
Moreover, a different $\gamma$/hadron separation can also be used.

For the purpose of this study, the energy range between 0.1\,TeV and 100\,TeV is divided into 30 equally-spaced logarithmic bins. 
Each of these energy bins consists of nine zenith-angle bins in the range from $0^\circ$ to $63^\circ$ but binned in $\cos{z}$. 
The binning with respect to $\theta$ is $0.25^\circ$ which is more than a factor of 2 of the energy-averaged $\gamma$-ray PSF of \hess\, \citep{Crab_paper}. 
Although the PSF for hadron-like events is larger than the one for $\gamma$-ray events, this is taken into account on average as hadron-like data from the FoV and the signal region are handled in the same way. 

The \hess\, analysis package (HAP; version \emph{12-03\_pl00}, DST version 12-03) was used for the event reconstruction \citep{Crab_paper}.

\subsection{Data processing}
For \hess\, data, the $\gamma$/hadron-separation cuts for $\gamma$-like events (\textit{standard cuts}) are $-2<\mathrm{MRSW}<0.9$ and $-2<\mathrm{MRSL}<2$, where MRSL is the analogously calculated mean reduced scaled length of an air-shower event \citep{Crab_paper}. 

In this work, the hadron-like regime is defined by $5<\mathrm{MRSW}<20$, an interval chosen to begin at a rather conservative value and also widened for the benefit of both a minimal gamma-contamination probability ($\sim5\,\sigma$ from $\gamma$-ray expectation) and higher event statistics.  
A selection with respect to the MRSL is not applied.

In the course of the data processing, events with larger reconstructed offsets than $\theta_\mathrm{max}=2^\circ$ are discarded to avoid camera-edge effects through truncated shower images (Sect.\,\ref{section:template}).\footnote{For comparison, the total \hess\, FoV has a radius of $\sim2.5^\circ$. Hence, for other IACT FoVs, $\theta_\mathrm{max}$ will be different.}
Additionally, only events within a maximum reconstructed distance of 1000\,m are considered. 
The energy threshold ${E}_\mathrm{thres}$ is conservatively set to be the energy bin following the one for which $E_\mathrm{bias}(z_\mathrm{obs},\omega_\mathrm{obs})\leq10\,\%$ is found in MC simulations. 
Here, $z_\mathrm{obs}$, $\omega_\mathrm{obs}$ are the mean values of zenith angle and wobble offset per observation run.
A typical value for $E_\mathrm{thres}$ is $\sim0.6\ldots0.7\,$TeV for \textit{Crab\,Nebula}-like observations ($z_\mathrm{obs}\approx50^\circ$, $\omega_\mathrm{obs}\approx0.5^\circ$).

All selection cuts applied to the \hess\, data are summarised in Table \ref{tab:datacuts}.

As discussed earlier in Sect.\,\ref{section:backgrounds}, known VHE $\gamma$-ray emission regions have to be excluded before the accumulation of the FoV data and the calculation of the background normalisation. 
These regions are excluded on the basis of the more sensitive TMVA analysis \citep[i.e. $\zeta$-std, ][]{TMVA}. 

\begin{table}
  \caption{\label{tab:datacuts} Summary of the event-wise \hess\, data selection for TBS used in this work.}
  \centering
    \begin{tabular}{l c c}
      \hline 
      \hline      
      Parameter & Gamma-like & Hadron-like\\
      \hline      \\[-2mm]     
      Energy (TeV) & $(0.1,100)$ & $(0.1,100)$\\
      MRSW & $(-2,0.9)$ & $(5,20)$\\
      MRSL & $(-2,2)$ & ---\\
      R$_\mathrm{max}$ (m) & 1000 & 1000\\
      $\theta_\mathrm{max}$ (deg) & 2 & 2\\
      $E_\mathrm{thres}$ & yes & yes\\
      \hline
    \end{tabular}
\end{table}

\subsection{Data sets}
To validate the method, known \hess\, sources with different spectral and morphological properties are investigated. 
Besides the standard procedures for obtaining high-quality data \citep[standard data quality selection in][]{Crab_paper}, the data in this work have been additionally selected according to the following aspects to assure a good performance of the IACT: the data sets are restricted to mean run zenith angles $z_\mathrm{obs} <55^\circ$. 
Only observations with all four telescopes in operation were considered. 
Observations with \hess\, are normally divided into runs of $\sim28\,$min of data taking, but are sometimes aborted due to clouds in the sky, for example. 
Hence, runs have to consist of at least 10\,min to be included in the analysis. 
Additionally, only runs not farther than $2.5^\circ$ away from the target position and which fully contain the source of interest are considered in order to avoid any bias when analysing extended sources. 
Moreover, these runs have to consist of data in the source region and as well as in the FoV (i.e. no \textit{On} runs and \textit{Off} runs). 
The selection criteria are summarised in Table \ref{tab:runcuts}.
\begin{table}
  \caption{\label{tab:runcuts} Summary of the run selection of \hess\, observations for TBS used in this work.}
  \centering
    \begin{tabular}{l c}
      \hline 
      \hline
      Parameter/Requirement & Value/Choice \\
      \hline\\[-2mm]
      \hess\, std quality selection$^{(1)}$ & yes\\
      \hline\\[-2mm]
      Number of active telescopes $N_\mathrm{Tel}$ & 4\\
      Run wobble offset $\omega_\mathrm{obs}$ (deg) & $<2.5$\\
      Run zenith angle $z_\mathrm{obs}$ (deg)& $<55$\\
      Live time (min) & $>10$\\ 
      \hline\\[-2mm]
      Full containment in FoV & yes \\
      Events from source region & yes\\
      Events from FoV & yes\\
      \hline
    \end{tabular}
     \tablebib{${(1)}$\,\citet{Crab_paper}.}
\end{table}

The \hess\, sources were selected on the basis of different aspects:
\begin{enumerate}
 \item[i.] source radius ($\theta_\mathrm{ON}=0.1^\circ\ldots1^\circ$), 
 \item[ii.] excess-to-background ratio (E/B $=11\ldots0.1$), and
 \item[iii.] source location (Galactic Plane and extra-Galactic).
\end{enumerate}
We note that the excess-to-background ratio is not an intrinsic source property, but more an inferred analysis-dependent quantity. 
The values stated here are the results of the standard template analyses in this work. 
On the basis of the data sets it is also possible to test the effect of different source properties/scenarios on the error-contribution terms (Eq. \ref{eqn:gamma_tbs_error}): 
\begin{enumerate}
 \item[a.] faint source (low E/B): $\sigma(N_\mathrm{g}^\mathrm{s})$,
 \item[b.] fewer FoV data: $(\sigma(\beta_\mathrm{eff})N_\mathrm{h}^\mathrm{s})^2$, and
 \item[c.] deep exposure: $(\sigma(\beta_\mathrm{eff})N_\mathrm{h}^\mathrm{s})^2$\,.
\end{enumerate}
In the following, the selected sources are briefly described. 
Their properties are summarised in Table \ref{tab:data_sets1} and \ref{tab:data_sets2}.
\begin{table}
  \caption{\label{tab:data_sets1} Properties of re-analysed H.E.S.S. data sets.}
  \centering
    \begin{tabular}{l c c c c }
      \hline 
      \hline      \\[-2mm]      
      Object & $\theta_\mathrm{ON}$ & $z_m$ $^{(1)}$ & $\omega_m$ $^{(2)}$ & $T$\\
      & deg & deg & deg & hrs\\
      \hline      \\[-2mm]
      Crab\,Nebula & 0.11 & 47 & 0.5 & 9.7\\
      \cena & 0.11 & 22 & 0.7 & 84.4\\
      HESS\,J1745--290 & 0.11 & 19 & 0.7 & 90.4\\
      HESS\,J1507--622 & 0.22 & 39 & 0.7 & 6.1\\
      Vela\,X & 0.8 & 25 & 0.9 & 54.6\\
      Vela\,Junior & 1.0 & 34 & 1.1& 12.1\\
      \hline
    \end{tabular}
    \tablefoot{	{${(1)}$ Median of the run-averaged zenith angles.} 
	        {${(2)}$ Median of the run-averaged wobble offsets.}}
\end{table}

\paragraph{The Crab\,Nebula :} The \crab\,\citep[][Data set III]{Crab_paper} is the strongest stable point-like source in the known TeV sky with a E/B $\gg1$. 
Hence, the TBS analysis of this source is rather considered to be a proof of concept and a test of the forward-folding techniques. 
The used run zenith-angle range is $z_\mathrm{obs}=45^\circ\ldots54^\circ$ with a wobble offset $\omega_\mathrm{obs}=0.5^\circ$. 
Compared to \citet{Crab_paper}, one less hour of data were used and essentially only runs with offsets higher than $0.5^\circ$ are missing here.

\paragraph{Centaurus\,A :} \cena\, \citep{CenA_paper} is an extra-Galactic point source, with a very low E/B and a low flux of less than $1\,\%$ of the \crab. 
In addition, this source is located in a clear FoV. 
The data set spans a run zenith-angle range of $19^\circ$ to $54^\circ$ and the offsets are between $0.4^\circ$ and $0.5^\circ$. 
This data set has 31 fewer hours of data, mainly due to the criteria of mandatory four-telescope runs and $z_\mathrm{obs}<55^\circ$.

\paragraph{HESS\,J1745--290 :} The Galactic Centre source \gc\, \citep{GC_paper} is a strong point-like source with E/B $\approx1$ and is located in a crowded FoV with many nearby sources and a complex diffuse $\gamma$-ray emission region. 
In addition, there are bright stars in the FoV. 
Three-telescope runs and especially runs with mean zenith angles above $55^\circ$ are not considered here but were used in \citet{GC_paper} to acquire more events at energies above 10\,TeV. 
To make up for the missing 30\,hrs, available data from the vicinty of \gc\, were used. The run zenith-angle range is $4^\circ$ to $54^\circ$ with an offset of $0.7^\circ$ to $2.0^\circ$.

\paragraph{HESS\,J1507--622 :} \hessjj\, \citep{J1507_paper} is an extended source located off-plane with an extent of $0.22^\circ$ in radius for the spectral reconstruction. 
The E/B is $0.3$. 
With only 6\,hrs of live time, this data set is about 3\,hrs smaller than the one used in \citet{J1507_paper}. 
The zenith-angle range is rather narrow ($38^\circ$ to $40^\circ$) and the run wobble offsets are within $0.6^\circ$ and $0.8^\circ$.

\paragraph{Vela\,X :} \vx\, \citep{VelaX_paper} is one of the closest pulsar-wind nebulae in the TeV sky with an extent of $0.8^\circ$ (called the \emph{cocoon}). 
It exhibits a very hard spectrum followed by a clear exponential cut off. 
In \citet{VelaX_paper}, the spectrum was reconstructed using the \textit{On}/\textit{Off} background. 
The data set is almost identical to the one used in \citet{VelaX_paper}, only some additional runs were included for which no \textit{Off} pairs were found in \citet{VelaX_paper}. The ranges are $z_\mathrm{obs}=22^\circ\ldots41^\circ$ and $\omega_\mathrm{obs}=0.1^\circ\ldots1.4^\circ$. 
\paragraph{Vela\,Junior :} \vjr\, \citep{VelaJr_paper} is a shell-type supernova remnant and with a diameter of $\sim2^\circ$ one of the largest sources in the TeV sky located in the vicinty of \vx. 
As for \vx, the spectrum was reconstructed using the \textit{On}/\textit{Off} background. The ranges are $z_\mathrm{obs}=25^\circ\ldots43^\circ$ and $\omega_\mathrm{obs}=0.5^\circ\ldots1.1^\circ$. 
In this work, 8 fewer hours of data were used than in \citet{VelaJr_paper}. 

\begin{table*}
  \caption{\label{tab:data_sets2} Results of the data analysis with the standard template background-model.}
  \centering
    \begin{tabular}{l c c c c c c c c}
      \hline 
      \hline      \\[-2mm]      
      Object & $N_\mathrm{g}^\mathrm{FoV}$ & $N_\mathrm{h}^\mathrm{FoV}$ & $\alpha_\mathrm{std}$ &$N_\mathrm{g}^\mathrm{s}$ & $N_\mathrm{h}^\mathrm{s}$ & $N_\mathrm{excess}$ & E/B & $S$ $^{(1)}$\\
      & & & & & & & & $\sigma$\\
      \hline      \\[-2mm]
      Crab\,Nebula & 81786 & 926331 & 0.088 & 3770 & 3629 & 3450 & 10.77 & 96.4 \\
      \cena & 707402 & 7835583 & 0.090 & 3115 & 30931 & 323 & 0.12 & 5.7 \\
      HESS\,J1745--290 & 302005 & 3175068 & 0.093 & 7245 & 38272 & 3670 & 1.03 & 50.8 \\
      HESS\,J1507--622 & 34668 & 407117 & 0.085 & 824 & 7694 & 169 & 0.26 & 6.1\\
      Vela\,X & 232820 & 2386719 & 0.098 & 114944 & 1081572 & 9439 & 0.09 & 27.3\\
      Vela\,Junior & 59026 & 640467 & 0.092 & 37033 & 353199 & 4482 & 0.14 & 23.2\\
      \hline
    \end{tabular}
    \tablefoot{	{${(1)}$ Significance calculated according to Eq. 17 of \citet{Lima}.}}
\end{table*}

\subsection{Results and discussion}

\begin{table*}
  \caption{\label{tab:spectra} Summary of TBS results and published \hess\, spectra.}
  \centering
    \begin{tabular}{l c c c c c c c c}
      \hline 
      \hline\\[-2mm]
      Object & Method & $E_\mathrm{min}$ & $E_\mathrm{max}$ & $\Phi_0(\mathrm{1\,TeV})$ & $\Gamma$ & $E_\mathrm{cut}$ & $\chi^2$/d.o.f.\\
       & & TeV & TeV &TeV$^{-1}$m$^{-2}$s$^{-1}$&  & TeV &  & \\
      \hline\\[-2mm]
      Crab Nebula & TBS & 0.63 & 20.0& $(3.82\pm0.11)\times10^{-7}$ & $2.40\pm0.06$ & $20.1 ^{+11.7}_{-5.4}$ & 7.24/12\\
   	  & Reflected regions$^{(1)}$ & 0.45 & 65 & $(3.84\pm 0.09)\times10^{-7}$ & $2.41\pm 0.04$ & $15.1 \pm 2.8$ & 12.6/9\\  
      \hline\\[-2mm]
      Centaurus\,A & TBS & 0.32 & 10.0 & $(2.10\pm 0.51)\times 10^{-9}$ & $2.74\pm 0.37$ & -- & 11.81/12 \\
                   & Reflected regions$^{(2)}$ & 0.25 & $\sim6$ & $(2.45\pm 0.52)\times 10^{-9}$ & $2.73\pm 0.45$ & -- & 2.76/4 \\		   
      \hline\\[-2mm]
      HESS\,J1745--290 & TBS & 0.2 & 15.9 & $(3.08\pm 0.15)\times10^{-8}$ & $2.10\pm 0.06$ & $13.0 ^{+7.1}_{-3.4}$ & 17.38/16\\
      (Galactic Centre)& Reflected regions$^{(3)}$ & 0.16 & 70 & $(2.55\pm 0.06)\times10^{-8}$ & $2.10\pm 0.04$ & $14.7 \pm3.4$ & 23/26\\			      
      \hline\\[-2mm]
      HESS\,J1507--622 & TBS & 0.5 & 6.3 & $(1.91\pm 0.41)\times 10^{-8}$ & $2.31\pm 0.30$ & -- & 6.69/9 \\
                       & Reflected regions$^{(4)}$ & $\sim0.5$ & 40.0 & $(1.8\pm 0.4)\times 10^{-8}$ & $2.24\pm 0.16$ & -- & --/4 \\	
      \hline\\[-2mm]
      Vela\,X & TBS & 0.79 & 63.1 & $(9.52\pm1.59)\times 10^{-8}$ & $1.32\pm0.29$ & $7.1 ^{+5.4}_{-2.1}$ & 6.68/15 \\
              & \textit{On/Off}$^{(5)}$ & 0.75 & $\sim60$& $(11.6\pm 0.6)\times10^{-8}$ & $1.36\pm 0.06$ & $13.9\pm1.6$ & --\\
      \hline\\[-2mm]
      Vela\,Junior & TBS & 0.3 & 20.0 & $(2.23\pm 0.30)\times 10^{-7}$ & $2.27\pm 0.13$ & -- & 11.23/16 \\
                   & \textit{On/Off}$^{(6)}$ & 0.3 & 20.0 & $(1.90\pm 0.08)\times 10^{-7}$ & $2.24\pm 0.04$ & -- & --\\		
      \hline
    \end{tabular}
    \tablefoot{The TBS energy ranges were chosen to match the \hess\, publication range. This was possible for \vx\, and \vjr. For \hessjj\, and \gc\, the full energy range in TBS (though limited by statistics) were used. The standard choice of the first energy bin, as used for the spectra of \cena\, and \crab, is the bin with the peak in the differential rate ($dN/dE$).}
    \tablebib{${(1)}$\,\citet{Crab_paper}; ${(2)}$\,\citet{CenA_paper}; ${(3)}$\,\citet{GC_paper}; ${(4)}$\,\citet{J1507_paper}; ${(5)}$\,\citet{VelaX_paper}; ${(6)}$\,\citet{VelaJr_paper}.}
\end{table*}

\begin{figure*}
  \centering
  \hbox{
    \includegraphics[width=0.33\textwidth]{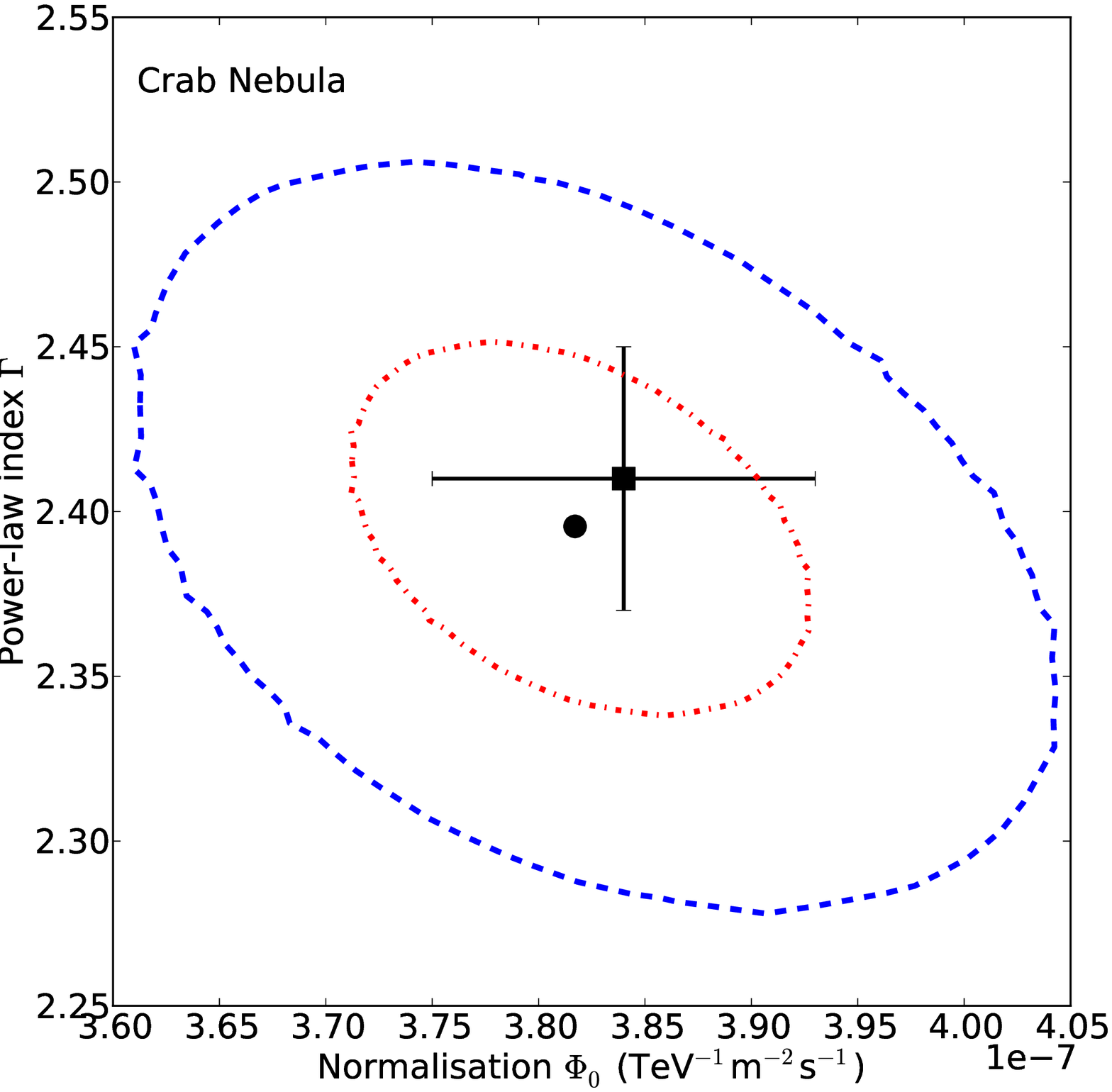}
    \includegraphics[width=0.33\textwidth]{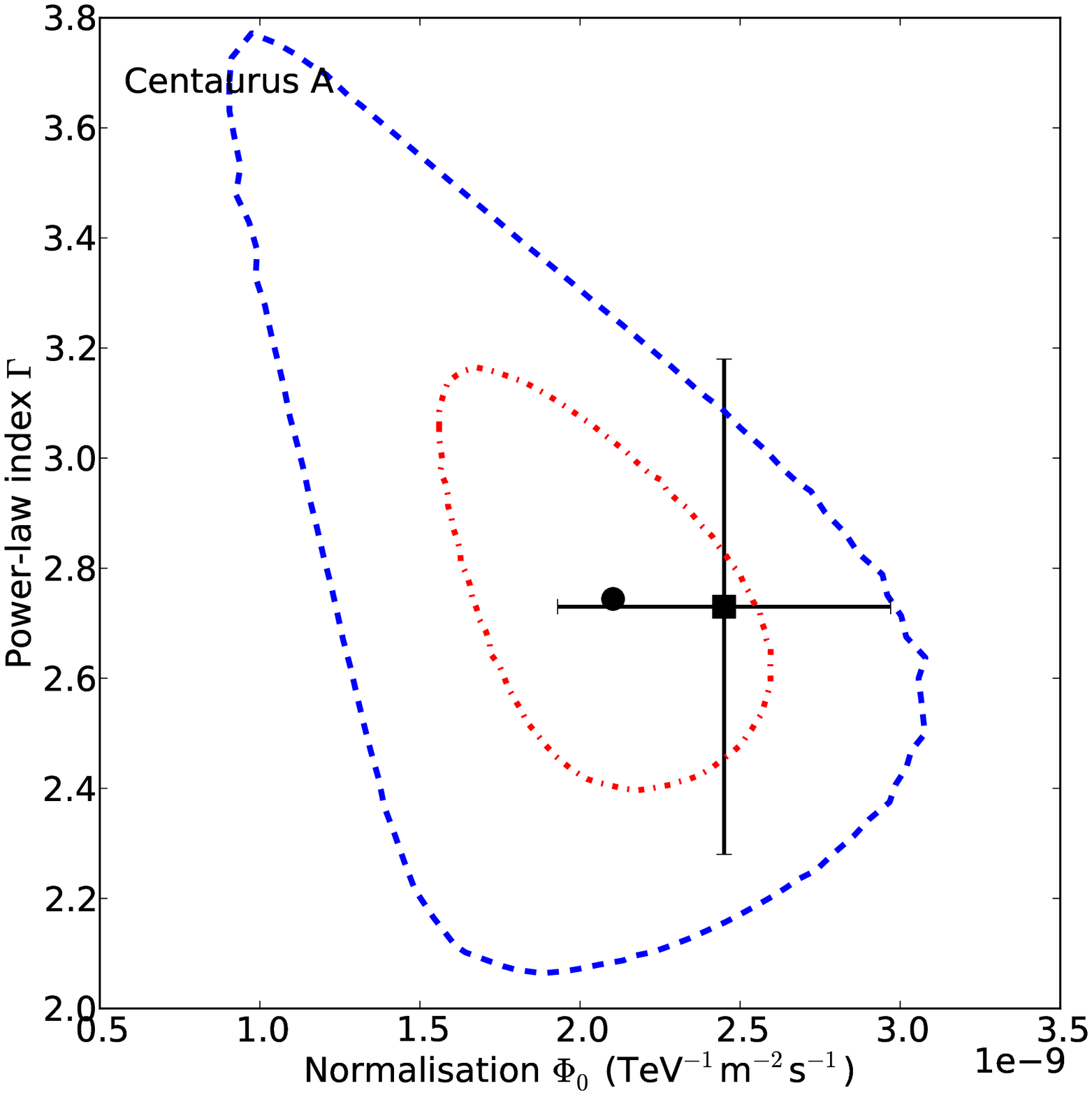}  
    \includegraphics[width=0.33\textwidth]{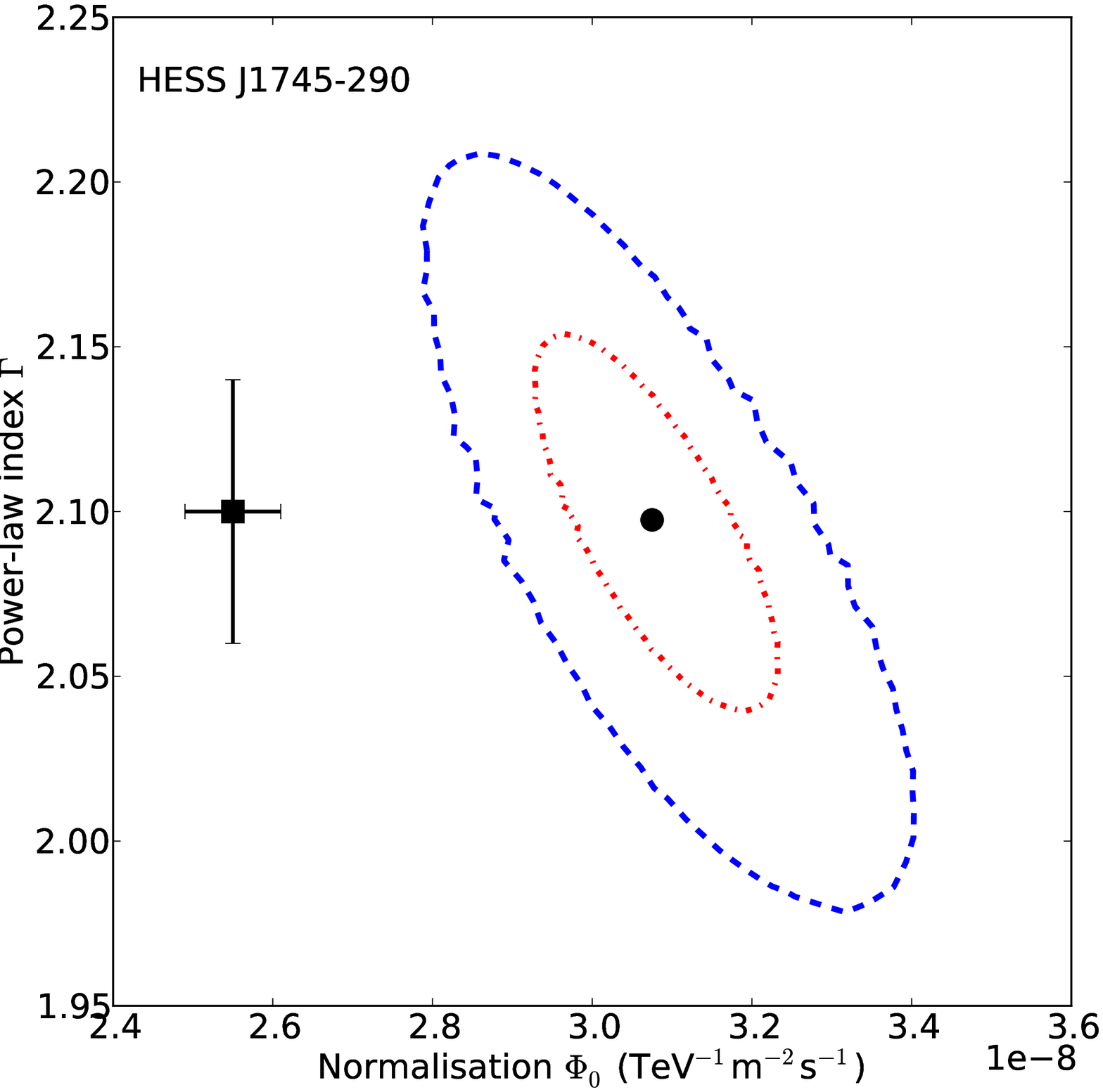}
  }
  \hbox{
    
    \includegraphics[width=0.33\textwidth]{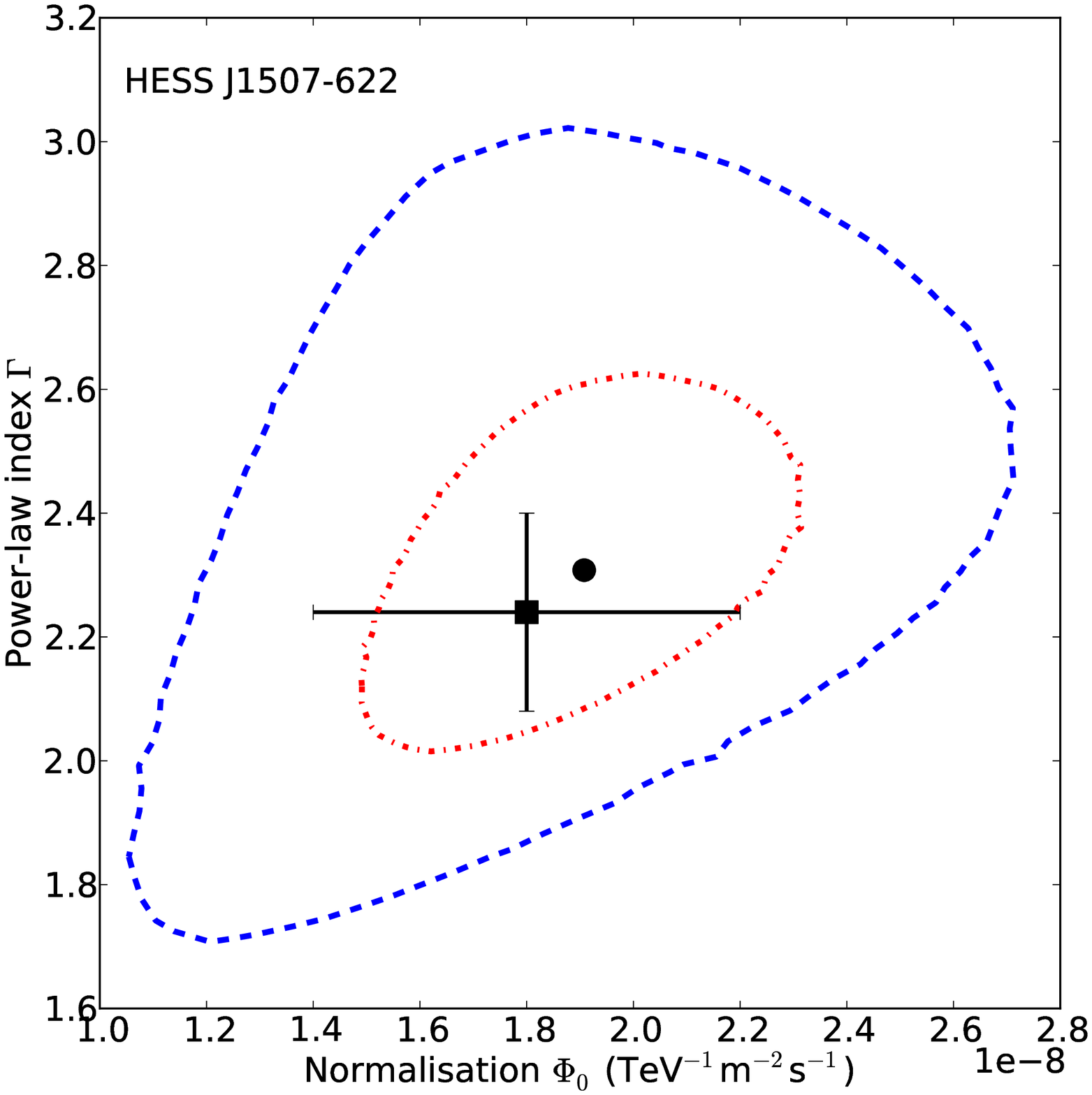}  
    \includegraphics[width=0.33\textwidth]{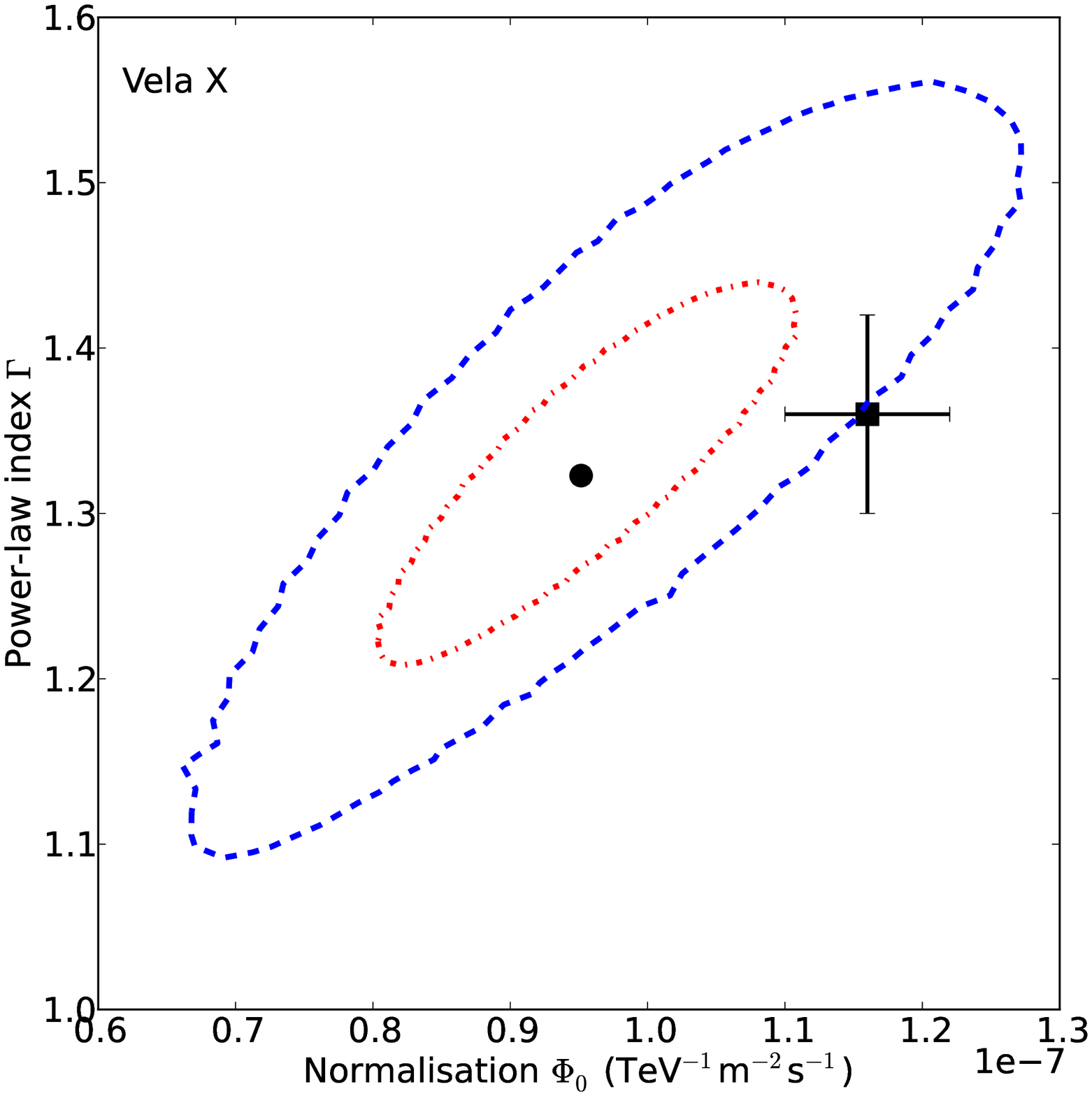}
    \includegraphics[width=0.33\textwidth]{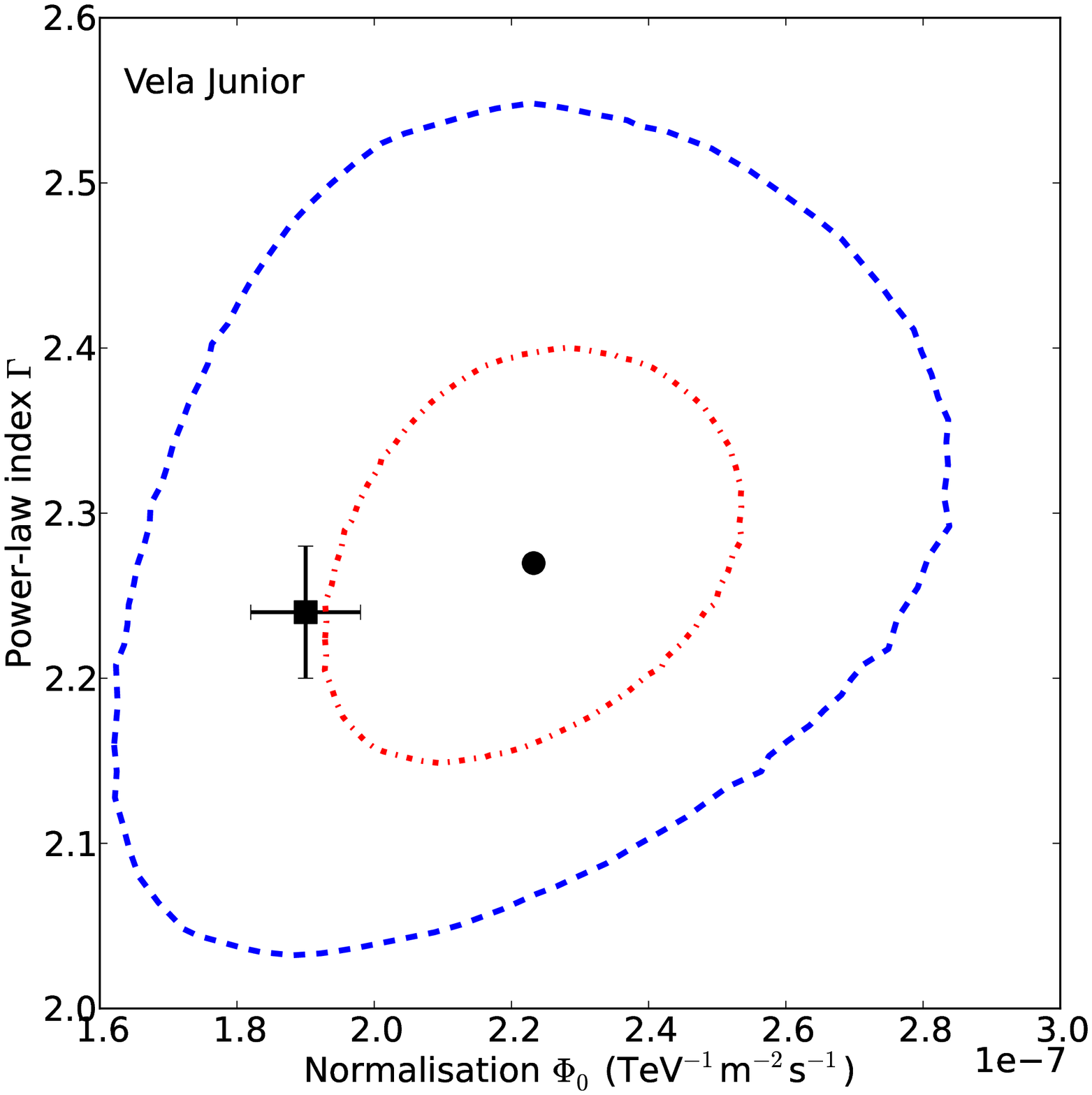}  
  }
  \caption{Correlation plots of the flux normalisation vs. the power-law index for the analysed sources. The inner red dash-dotted line represents the 1\,$\sigma$ contour, whereas the blue dashed line indicates the 2\,$\sigma$ contour. The black dot is the best fit from the TBS analysis. The square with its error bars marks the published \hess\, result.}
  \label{fig:corrplots}
\end{figure*} 

In general, the results of TBS are compatible with the published \hess\, spectra. 
The flux normalisations and the power-law indices are mostly within 1\,$\sigma$ with respect to the published results (see Table \ref{tab:spectra} and Fig.\,\ref{fig:corrplots}). 
The spectral shape used to determine the best-fit spectrum (either a simple power or a power law with exponential cut off) is taken from the respective \hess\, publication.
The differential energy spectra of the \crab\, from \citet{Crab_paper} and from this work are visually compared in Appendix\,\ref{crab_spectrum}.
Spectra of the other sources are subject to ongoing \hess\, efforts (source studies and systematic studies at energies above $\sim10$\,TeV).
Therefore, flux points of these sources derived with TBS cannot be shown. 
The compatibility of the TBS results with the other standard methods to reconstruct spetra is demonstrated in Table \ref{tab:spectra} and Fig.\,\ref{fig:corrplots}.

In the following, the results are discussed source-by-source.

\paragraph{The Crab\,Nebula :} Both data sets and the spectral results are compatible. 
Up to $\sim12$\,TeV, the template correction $\beta_i$ is calculated mainly through interpolation. 
At higher energies, the sample of $\beta_i$ are calculated through extrapolation towards lower offsets. 
Above 20\,TeV in combination with the binning in energy, statistics from the signal region are not sufficient. 
The statistical errors on the flux normalisation and on the power-law index are nearly identical.

\paragraph{Centaurus\,A :} The data set analysed in this work consists of 31 fewer hours of live time compared to the \hess\, publication and yet, the TBS spectral results are compatible with the \hess\, publication. 
The spectrum extends to higher energies (up to 10\,TeV) than the \hess\, spectrum, as here a forward-folding technique is used to determine the best-fit spectrum with more degrees of freedom.
The usage of 12 degrees of freedom in the foward folding led to a smaller statistical uncertainty on the power-law index. 
As for the \crab\, data, $\beta_i$ are mainly calculated via interpolation.

\paragraph{HESS\,J1745--290 :} \gc\, is the source with the largest deviation in the normalisation between \citet{GC_paper} and this work, although the power-law indices and the cut-off energy are nearly identical. 
This could be explainable as the high zenith-angle observations are not included in our sample (which makes the spectral reach of TBS end at 16\,TeV ) and that $\sim30$\,\% additional data in the FoV have been used to bring the live time to $\sim90\,$hrs, or perhaps it is due to much larger exclusion regions. 

In general, for this complex region, any difference in the treatment of the diffuse emission is a possible explanation for the observed difference. 
Because of insufficient statistics in the signal region, the TBS spectrum extends only up to 16\,TeV. 
From $\sim8\,$TeV onwards, an increasing fraction of $\beta_i$ are calculated by extrapolating the template correction to lower offsets.

However, given the deep exposure, these differences could also be perhaps due to the systematics of the different analysis procedures.
Hence, a $\sim20\,\%$ difference as seen here in the normalisation may not be a major concern for this source in its complicated FoV. 
As a test, the large data set was split in half to check for any variation affecting the TBS correction. 
However, the results were compatible with the findings presented here.

\paragraph{HESS\,J1507--622 :} Although about $\sim30\,$\% fewer data are used for this source in this work, the results are compatible within 1\,$\sigma$. 
The values for $\beta_i$ were mainly calculated through interpolation. With only 6\,hrs of data, the analysis of this source suffers from low statistics in $N_\mathrm{g}^\mathrm{s}$ towards higher energies and the spectrum extends only up to 6\,TeV. 

\paragraph{Vela\,X :} Updated results on \vx\, have been published by the \hess\, collaboration. 
With more than 50\,hrs of data, sufficient statistics are provided to extend the spectrum up to 60\,TeV. 
Up to 10\,TeV, the $\beta_i$ are determined via interpolation; at higher energies, nearly half of the $\beta_i$ quantities are also calculated through extrapolation of $\alpha$ to lower offsets. 
At energies above 25\,TeV, the $\beta_i$ are calculated mainly via extrapolation towards lower offsets. 
For this source, the normalisation estimated with TBS is 20\,\% (or almost 2\,$\sigma$) lower than the published \hess\, result. 
Although the index is reproduced, the spectral cut off is at a lower energy. 
This may be due to systematic differences between the template background in TBS and the \emph{On}/\emph{Off} background (as \hess\, observations are not conducted in an \emph{On}/\emph{Off} scheme). 
As a test for a systematic effect of the correction within TBS, the data set was reduced by half and re-analysed. 
However, the normalisation of the resulting spectrum was still found to be 1.5\,$\sigma$ lower than in \citet{VelaX_paper} and compatible with the previous TBS result. 
From $\sim13\,$TeV onwards, the $\beta_i$ are calculated to a dominant fraction from extrapolation towards lower offsets. 

\paragraph{Vela\,Junior :} The results of TBS for \vjr\, are compatible with \citet{VelaJr_paper}. 
The fraction of $\beta_i$ that are estimated through extrapolation to lower offsets is from lower energies on to 2\,TeV at a 10\,\% level and constantly increases up to 80\,\% at 20\,TeV.

\subsection{Errors and limitations} 
The normalisation $\alpha$ has a non-negligible error unlike in other background estimation methods. 
Moreover, the corrected quantities (signal hadrons) are statistically not independent within a specific bin in ($E$,$z$,$\theta$). 
Because of these circumstances, higher errors on $N_\mathrm{excess}$ than in other methods are found. 
As long as the statistical errors on $\alpha$ and therefore on $\beta_\mathrm{eff}$ are smaller than those from the analysed source ($N_\mathrm{g,h}$), systematic effects will not have an impact on the correction. 
To investigate this, Eq. \ref{eqn:gamma_tbs_error} is re-written to resemble a sum of variances:
\begin{equation}
 \mathrm{Var}\left(N_\mathrm{excess}\right) = \mathrm{Var}\left(N_\mathrm{g}^\mathrm{s}\right)+\mathrm{Var}\left(\beta_\mathrm{eff}^2N_\mathrm{h}^\mathrm{s}\right)+\mathrm{Var}\left(\left(\sigma(\beta_\mathrm{eff})N_\mathrm{h}^\mathrm{s}\right)^2\right)\,.
\end{equation}
For legibility reasons, we write
\begin{equation}\label{eqn:var}
 \mathrm{Var}\left(\mathrm{e}\right) = \mathrm{Var}\left(\mathrm{g}\right)+\mathrm{Var}\left(\mathrm{h}\right)+\mathrm{Var}\left(\beta_\mathrm{eff}\right) \,.
\end{equation}
These terms are the variance contribution as a sum of statistical errors on $N_\mathrm{g}^\mathrm{s}$, on $N_\mathrm{h}^\mathrm{s}$, and on the effective template correction $\beta_\mathrm{eff}$. 
However, $\mathrm{Var}\left(\beta_\mathrm{eff}\right)$ also includes systematic errors (see Appendix \ref{appendix:tbs}). 
In Fig.\,\ref{fig:varplots}, the contribution of these three components to the total excess error $\mathrm{Var}\left(\mathrm{e}\right)$ per energy per source is shown. 
For all sources but \vx\, and \vjr\, the uncertainties related to $\beta_\mathrm{eff}$ are negligible as the dominant contribution to the overall statistical errors arises from $N_\mathrm{g}^\mathrm{s}$.
For the large sources \vx\, and \vjr, the uncertainty on $\beta_\mathrm{eff}$ is on average around 0.9 of the total excess error.

Based on the results, a low E/B source does not constitute a problem for TBS \textit{per se} as the \cena\, spectrum is sufficiently well reproduced. 
In addition, the results for sources with a deep exposure (\gc, but also \cena) are consistent with the \hess\, measurement. 
However, for low E/B and very extended sources like \vx\, and \vjr\, one may approach the limits of TBS. 
Large sources naturally leave comparatively fewer data in the FoV to calculate the background normalisation (but on the other hand exhibit very low relative errors on the gamma-like and hadron-like counts).
On average, the correction has to be calculated through extrapolation to a much higher fraction than in other cases. 
The extrapolation has a much larger uncertainty (no \textit{a priori} knowledge how $\alpha$ behaves for a certain ($E$,$z$,$\theta$) bin). 
This constitutes the largest uncertainty in TBS when calculating the template correction in ($E$,$z$,$\theta$) and the effect is stronger the larger the source is. 

In general, the E/B ratio depends on the performance of the $\gamma$/hadron separation.
Hence, when using a more sophisticated method as outlined in Sect. \ref{section:intro}, a higher E/B ratio compared to the simple MRSW selection is expected.

The \textit{a priori} choice of binning in energy, zenith angle, and camera offset might not be optimal. 
However, the analyses with different binnings in energy (5 or 20 logarithmic bins per decade in energy) and zenith angle (a factor of two finer binning) were compatible within $\sim1\,\sigma$ with respect to the results presented here. 
In general, a too fine binning in any of these parameters (i.e. energy, zenith angle, and offset) will lead to a higher fraction of extrapolation and hence to higher uncertainties. 
In addition, a different and more gamma-like MRSW selection ($0.9 < \mathrm{MRSW} < 5$) for the hadron-like regime did not result in a significant change with respect to the results presented here. 
However, the uncertainties on $\beta_\mathrm{eff}$ increase as fewer data are available compared to the larger MRSW selection ($5< \mathrm{MRSW}<20$).

\subsection{$\alpha_\mathrm{std}$ vs. $\alpha(E)$ vs. $\beta_\mathrm{eff}$} 
\begin{figure*}[t]
  \resizebox{\hsize}{!}{
    \includegraphics{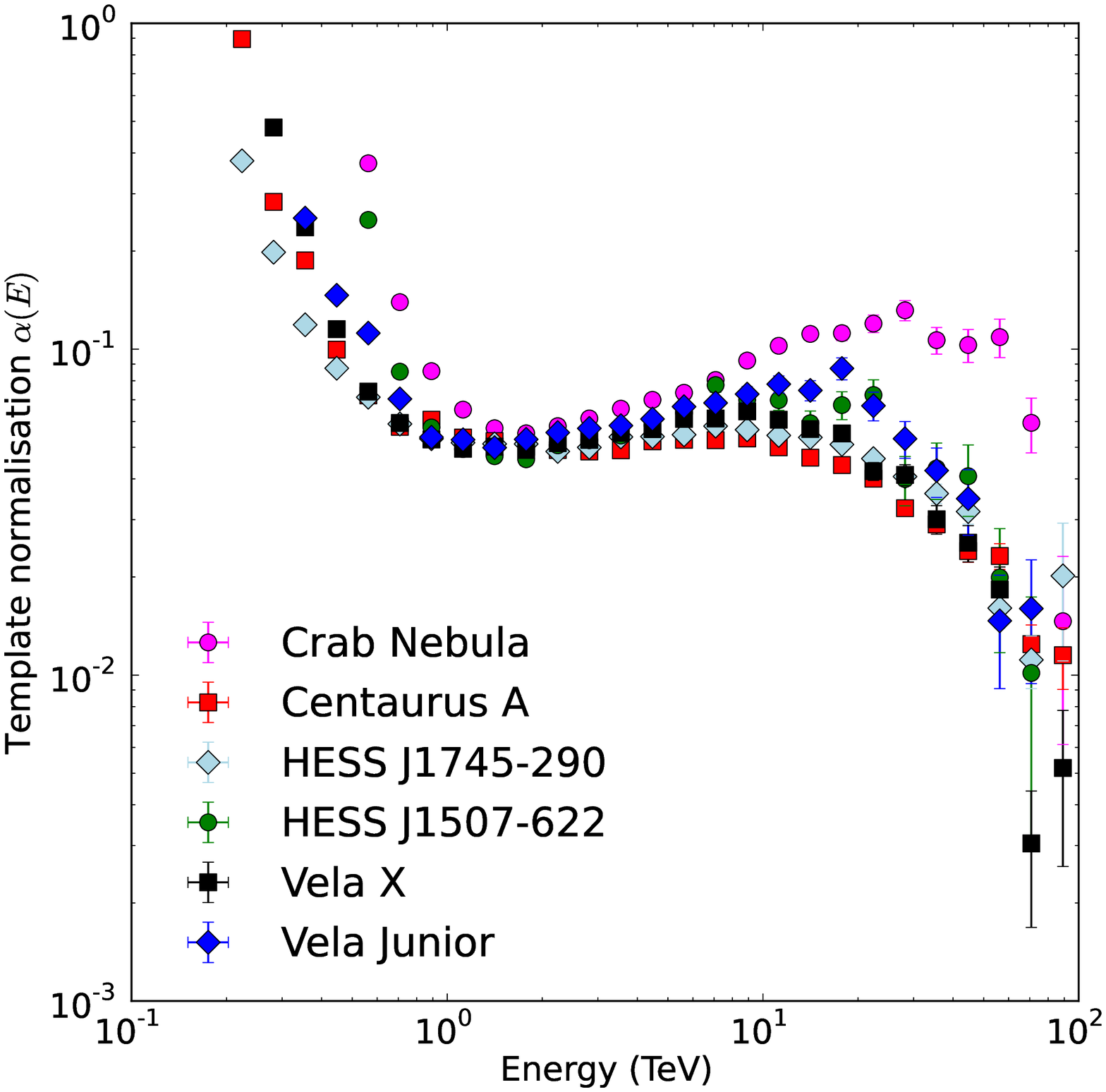}
    \includegraphics{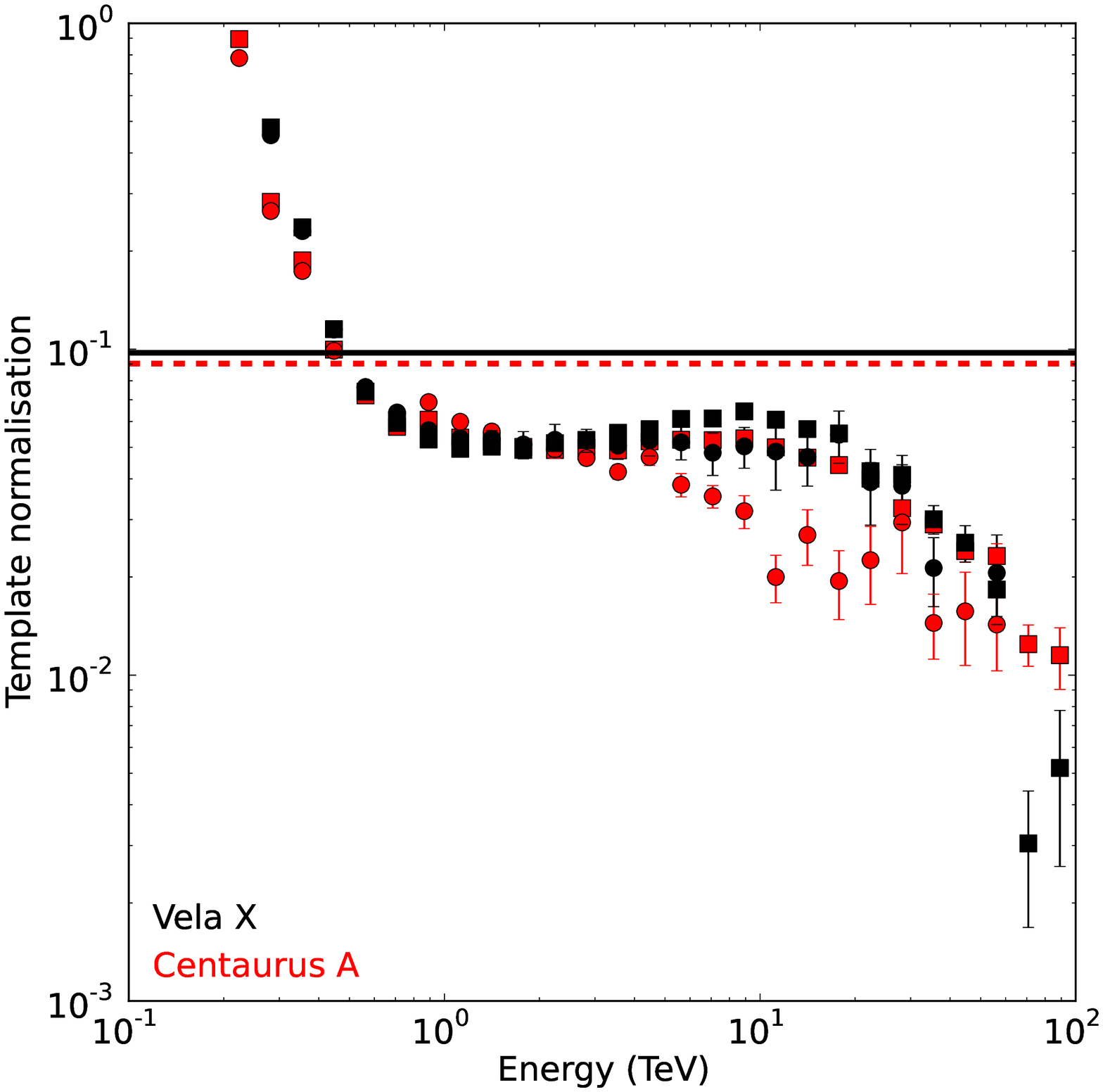}}
  \caption{TBS and template correction. \emph{Left:} $\theta$ and $z$-integrated template correction $\alpha(E)$ from data on the sources analysed in this work. \emph{Right:} Effective template correction $\beta_\mathrm{eff}$ (circles) and $\alpha(E)$ (squares) for the sources \cena\, (red) and \vx\, (black). The black solid line represents the integrated value $\alpha_\mathrm{std}$ of \vx\, whereas the red dashed line represents the overall template correction from the \cena\, data set. See text for further information.}
  \label{fig:tbs_beta_eff_vs_alpha}
\end{figure*}
In Fig.\,\ref{fig:tbs_beta_eff_vs_alpha} (left), the zenith-angle and offset-integrated standard template normalisation $\alpha(E)$ is shown for all sources analysed in this work. 
In general, they all follow the same trend. 
Differences at lower energies are mainly due to different zenith angles and therefore different energy thresholds at which data are accumulated. 
At higher energies, the effective area drops because of selection and quality cuts corresponding to lower values of $\alpha(E)$. 
However, for data sets consisting of a large fraction of high zenith-angle observations (the effective area drops later), comparatively more gamma-like events from the FoV data are found leading to rise in $\alpha(E)$, e.g. to be seen for the \crab\, data set.

The different normalisations $\alpha(E)$ could be an indication that the background does not behave in the same way from certain energies onwards for specific FoVs.\footnote{A fact that could be of importance when considering the production of a complete set of $\alpha(E,z,\theta)$ from \emph{Off}/extra-Galactic data.} 
However, these differences are also a consequence of the different zenith angles used (see Table \ref{tab:data_sets2}).

On the right-hand side, $\alpha_\mathrm{std}$ vs. $\alpha(E)$ vs. $\beta_\mathrm{eff}$ are shown for the analysis \cena\, and \vx. 
It can be clearly seen that the overall normalisation $\alpha_\mathrm{std}$ cannot be used for the reconstruction of energy spectra.
The quantities $\alpha(E)$ and $\beta_\mathrm{eff}$ behave similarly up to energies of $\sim3$\,TeV. 
Especially, for \cena\, the deviation between $\alpha(E)$ and $\beta_\mathrm{eff}$ is clearly seen at higher energies. 
Hence, a pure energy-dependent treatment of the background normalisation would not lead to the correct background estimate in a spectral reconstruction.

\section{Summary and conclusions}\label{section:summary}
In this work, we proposed and tested a new method for estimating the background for VHE $\gamma$-ray spectra. 
The Template Background Spectrum (TBS) enables spectral reconstruction in crowded FoVs and for extended sources where other standard methods fail mainly because of geometrical limitations. 
Moreoever, no additional \textit{Off} data have to be used. 

The basic idea is the accumulation of data binned in energy, zenith angle, and camera offset to create template-correction lookups from the FoV data which are then used to correct the identically binned data from the source of interest. 
This means that, the problem of low statistics in the run-by-run analysis can be circumvented.
Moreover, the template correction accounts for any differences in the gamma-like and hadron-like regimes as every bin in $(E,z,\theta)$ is already corrected, for example with respect to the exposure or the bin size, so even the intrinsic differences between the gamma-like (for which the MC data are made) and the hadron-like regime, for example the effective area, are on average accounted for.

Template Background Spectrum was tested on published H.E.S.S. data from various FoVs and different types of sources (strong to faint, point-like to extended, located in rather empty or in rather crowded FoVs). On average, good agreement was found between the spectra reported by the \hess\, collaboration and in this work. 
Although \hess\, data were used, the method is useable for any IACT data. 

In a future effort, the studied parameter range will be extended to increase the performance of TBS, especially at the highest energies.
On the basis of the \hess\, data this means primarily including MC data of large zenith angles ($>55^\circ$) and also two-to-three-telescope observations to the set of instrument response files.

Unlike other background-estimation methods, TBS does not require symmetric ON/signal regions which makes it possible to define arbitrarily formed (i.e. better suited) regions for spectral analyses.
Now, besides the \textit{On}/\textit{Off}-background method and the reflected-region background, TBS is the third general method for reconstructing energy spectra. 
Compared to the two former ones, new and different problems have to be tackled which can lead to higher statistical and systematic errors on the background normalisation.

The ansatz of an energy and zenith-angle-dependent treatment of the template normalisation offers new possibilities. 
For example, it enhances the power of the normal template background method in producing skymaps in which every event would then be corrected according to its ($E$,$z$,$\theta$) dependence, and therefore contribute to morphological studies of (extended) sources in which the correction to the data is done in ($E$,$z$,$\theta$) space. 

Using the latest advances in background rejection, the performance of TBS could be further enhanced. 
In general, any selection cut or algorithm that improves the separation of gamma-like and hadron-like events compared to the Hillas approach will possibly lead to an improvement of TBS. 
In addition, as a further improvement, interpolation techniques other than linearly interpolating may be investigated. 

The main uncertainty of TBS is the extrapolation. 
To enhance the performance of TBS, the accuracy of the extrapolation has to be increased. 
Although camera-acceptance lookups are in use when generating skymaps, thus avoiding extrapolation issues in complex FoVs or for large sources, it has to be investigated if lookup tables generated from extra-Galactic observations and split into in $E,z,\theta$ can be applied to any FoV and source, or if the systematic uncertainties in the background, of the template background method, or of TBS dominate. 
Most probably, one would (at least) introduce a time dependence $t$ as a new parameter of TBS to calculate $\alpha(E,z,\theta,t)$. However, to assure sufficient statistics at higher energies, an increasing bin width at energies beyond $\sim5$\,TeV may also suffice.
Currently, sources like \vx\, or \vjr\, fill a large fraction of the FoV leaving relatively less data to calculate the background normalisation.
If the next generation of IACTs is equipped with larger FoVs ($>5^\circ$), more data would be available to calculate $\alpha$ and thus reduce uncertainties introduced by the extrapolation of the template correction.

In principle, the same approach can be used with the ring background method \citep{BgTechs} which is only used for skymap generation. 
Here a ring is placed around the ON region to determine the background. 
Similar to TBS, one would have to compute the background normalisation in energy, zenith angle, and camera-offset space.
However, for large sources and in crowded regions its application is more strongly limited than the template background model and TBS.
Its advantage is that only the gamma-like sample is used and therefore the uncertainties related to the hadron-like background do not apply here.

This concept of an energy-dependent treatment of a non-spatial parameter to estimate the background contribution to energy spectra could in principle also be applied to other air-shower experiments than IACTs.

In \citet{Fernandes2014}, a likelihood estimation method similar to that used in \citet{logLike} has been tested and found to reduce the uncertainty on the best-fit spectral parameters in the forward folding, especially for extended sources.

\begin{appendix}
\section{Estimating the TBS correction}\label{appendix:tbs}
After creating the lookup of $\alpha(\Delta{E},\Delta{z},\Delta\theta)$ in bins of energy $\Delta{E}$, zenith angle $\Delta{z}$, and camera offset $\Delta\theta$, the task is to correct each hadron-like event from the signal region with its reconstructed energy $E_i$, zenith angle $z_i$ and offset from the camera $\theta_i$.

Rather than simply using the binned $\alpha(\Delta{E},\Delta{z},\Delta\theta)$ as the correction factor to the data, the template correction is determined through interpolation and extrapolation of $\alpha$ values.
Technically speaking, the $\alpha$ values are considered the nodes of interpolation and extrapolation and on this basis the event-specific correction factor $\beta_i$ is calculated.
In this work, a simple linear-interpolation procedure is used to determine the individual $\beta_i$ factors.
In the following, the linear interpolation and the extrapolation tasks as well as the error propagation of the TBS correction are described. 

The aforementioned nodes are prone to statistical fluctuations. 
In addition, the choice of the binning may introduce systematic effects in the correction. 
Any uncertainty in the interpolation or extrapolation will likely propagate into the final spectrum. 
Its effect is stronger the more often an improperly determined $\alpha$ is used to correct the data. 
For the error propagation, it has to be accounted for that interpolated and extrapolated values within the same $(E,z,\theta)$ bin are mostly not statistically independent.

\subsection{Interpolation and error estimation}\label{appendix:interpolation}
\begin{figure}[t]
  \centering
  \resizebox{0.9\hsize}{!}{
   \includegraphics{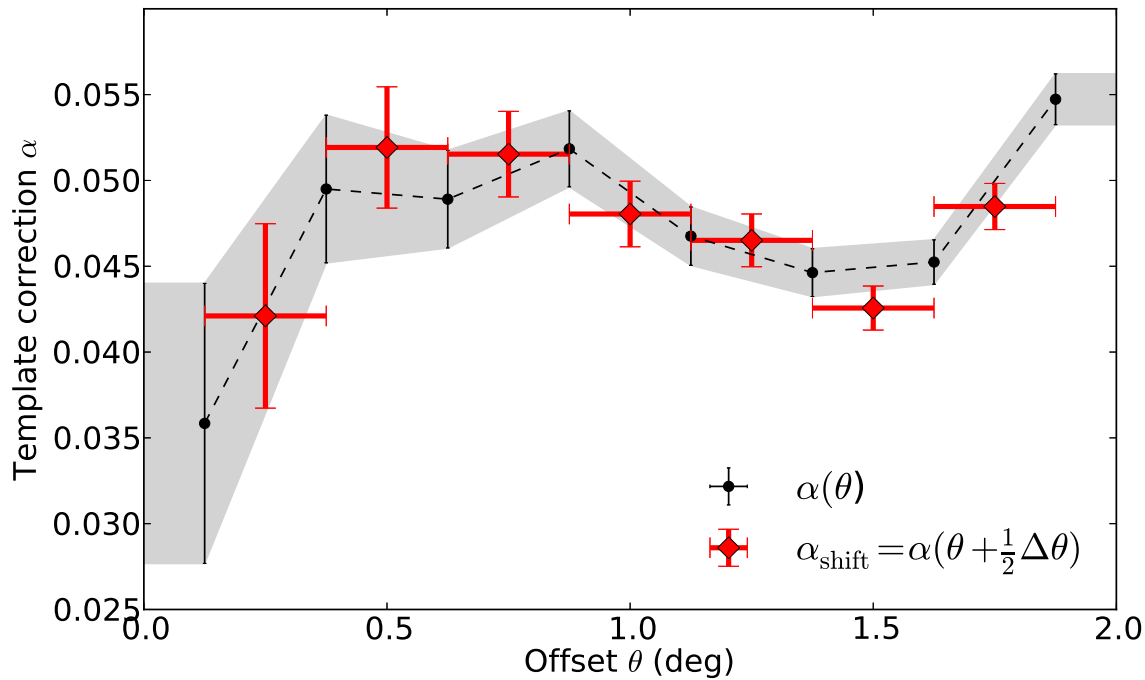}}
  \caption{Example of interpolation of $\alpha(\theta)$ for TBS based on \hess\, data on \vx\, \citep{VelaX_paper}. The $\alpha$ nodes calculated according to Eq. \ref{eqn:tbs_alpha} are shown as black circles together with the corresponding $1\,\sigma$ envelope (grey-shaded area). The dashed black line illustrates the interpolation line along which a $\beta_i$ value will be obtained. The parameter $\Delta\theta$ is the bin width in $\theta$. The shifted nodes $\alpha_\mathrm{shift}$ are drawn as red diamond markers. See text for further information.}
  \label{fig:tbs_interpolation}
\end{figure}
In Fig.\,\ref{fig:tbs_interpolation}, the $\alpha(\theta)$ is plotted for a fixed energy and zenith-angle bin. 
At first look, the assumption of a linear dependence of $\alpha(\theta)$ (black circles) is valid within statistical errors between two neighbouring $\alpha$ nodes. 
This is to be seen when one compares the nodes with those computed with a shift of half a bin width in $\theta$ (red diamonds). 
In most cases, this shifted set of nodes $\alpha_\mathrm{shift}(\theta)$ lies within $1\,\sigma$ of the expected linear connection between the $\alpha(\theta)$ (compare the red diamonds with the dashed line indicating the interpolation). 
However, if $\alpha_\mathrm{shift}(\theta)$ exhibits a larger deviation from the assumed linear connection (dashed line) this is an indication that the simple linear interpolation may not be valid for this bin.
Not accounting for these outliers will bias the correction and lead to an overestimate or underestimate of the correction.
In the following, the linear interpolation and its respective error, also accounting for the above mentioned outliers, is described.

For a hadron-like event $i$ from the signal region and a complete set of $\alpha$ nodes for a fixed $({\Delta}E,{\Delta}z)$, the interpolated quantity is
\begin{equation}\label{eqn:beta_interpolate}
 \beta_i(\theta_i) = \frac{\theta_\mathrm{high}-\theta_i}{\theta_\mathrm{high}-\theta_\mathrm{low}}\alpha(\theta_\mathrm{low})+\frac{\theta_i-\theta_\mathrm{low}}{\theta_\mathrm{high}-\theta_\mathrm{low}}\alpha(\theta_\mathrm{high})
\end{equation}
\begin{equation}\label{eqn:beta_interpolate2}
 \beta_i(\theta_i) = c_\mathrm{low}\alpha_\mathrm{low}+c_\mathrm{high}\alpha_\mathrm{high},
\end{equation}
with $c_\mathrm{low}+c_\mathrm{high}=1$ and where $\theta_i$ is the specific event offset with its next neighbours $\theta_\mathrm{low},\theta_\mathrm{high}$ and their corresponding nodes $\alpha_\mathrm{low},\alpha_\mathrm{high}$. 
$\alpha_\mathrm{low},\alpha_\mathrm{high}$ are statistically independent.
In principle, one could calculate the respective error on $\beta_i$ through simple error propagation, but his would not account for the systematic uncertainty discussed above and would underestimate the errors.

To estimate the interpolation error, two quantities are calculated. 
First, the $1\,\sigma$ envelope $\epsilon$ around the $\alpha$ nodes is calculated (grey-shaded area in Fig.\,\ref{fig:tbs_interpolation}):
\begin{equation}
\epsilon(\theta_i)=c_\mathrm{low}\sigma(\alpha_\mathrm{low})+c_\mathrm{high}\sigma(\alpha_\mathrm{high})\,.
\end{equation}
Second, to estimate the uncertainty introduced through the \textit{a\,priori} binning, a second set of nodes $\alpha_\mathrm{shift}$, shifted by $0.5\Delta\theta$ is calculated. 
Whenever $\alpha_\mathrm{shift}$ does not lie within $\epsilon$, the distance $\delta=\left|0.5(\alpha_\mathrm{low}+\alpha_\mathrm{high})-\alpha_\mathrm{shift}\right|$ is used as the error estimate:
\begin{equation}\label{eqn:beta_interpolate_error}
 \sigma(\beta_i) = \max(\epsilon,\delta)\,.
\end{equation}
The shifted nodes $\alpha_\mathrm{shift}$ are mostly within $\epsilon$ which shows that the binning chosen is stable in $\theta$. 
However, outliers used in the interpolation can lead to residuals in the spectrum if a large fraction of hadrons does fall into this bin (e.g. in Fig.\,\ref{fig:tbs_interpolation} for $\theta=1.5^\circ$).

\subsection{Extrapolation and error estimation}\label{appendix:extrapolation}
\begin{figure}[t]
  \centering
  \resizebox{0.9\hsize}{!}{
    \includegraphics{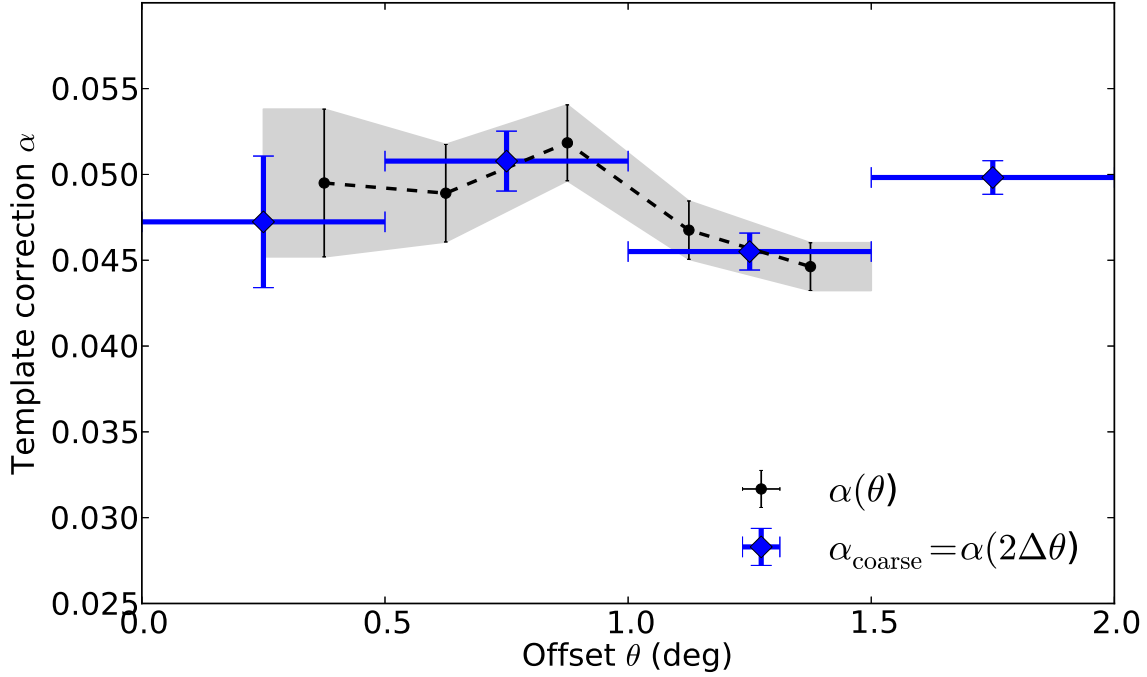}}
  \caption{Example of an extrapolation of $\alpha(\theta)$ for TBS based on the same data as in Fig.\,\ref{fig:tbs_interpolation}, but now with additional next-coarser binned nodes $\alpha_\mathrm{coarse}$ drawn (blue diamonds) which are used to extrapolate to lower or higher offsets. See text for further information.}
  \label{fig:tbs_extrapolation}
\end{figure}
There are two types of extrapolation: 
i) the extrapolation within $({\Delta}E,{\Delta}z,{\Delta}\theta)$ when there is not enough data to calculate $\alpha$ for the complete $\theta$ range within $({\Delta}E,{\Delta}z)$ and ii) the extrapolation for $({\Delta}E,{\Delta}z)$ when no $\alpha$ node could be calculated for a fixed $({\Delta}E,{\Delta}z)$ bin (see Fig.\,\ref{fig:tbs_extrapolation}).

For the first case, again, different sets of nodes $\alpha_\mathrm{coarse}$ are computed for coarser binnings in $\theta$, each binning increased by factor of 2. 
So, if $\theta_i$ is within the next closest bin for which a regularly computed correction is available, its value $\alpha_\mathrm{next}$ and error $\sigma(\alpha_\mathrm{next})$ are taken as the correction estimate:
\begin{equation}\label{eqn:beta_extrapolate}
 \beta_i(\theta_i) = \alpha_\mathrm{next}\,\mathrm{ with\, error: }\,\sigma(\beta_i)=\sigma(\alpha_\mathrm{next})\,.
\end{equation}
If the requested $\theta_i$ is farther away, the next coarser binning with respect to to the normal set of $\alpha$ is used, the TBS correction is then conservatively estimated to be their average: 
\begin{equation}\label{eqn:beta_extrapolate1}
 \beta_i(\theta_i) = 0.5(\alpha_\mathrm{next}+\alpha_\mathrm{coarse})\,.
\end{equation}
In this case, $\alpha_\mathrm{next}$ and $\alpha_\mathrm{coarse}$ are mostly statistically dependent. The error is then estimated on the basis of the relative error:
\begin{equation}\label{eqn:beta_extrapolate_error}
 \sigma(\beta_i) = \beta_i \frac{\sigma(\alpha_\mathrm{next})}{\alpha_\mathrm{next}}\,.
\end{equation}
If even coarser binnings have to be used, the maximum relative error of the regular $\alpha$ sample is used:
\begin{equation}\label{eqn:beta_extrapolate_error2}
 \sigma(\beta_i) = \beta_i\max\left(\frac{\sigma(\alpha)}{\alpha}\right)\,.
\end{equation}
In Fig.\,\ref{fig:tbs_extrapolation}, without the knowledge of the coarser binning, $\alpha(\theta>1.4^\circ)$ would be interpolated to lower values as this would seemingly be the trend of the data. 
However, $\alpha$ increases instead (see Figs.\,\ref{fig:tbs_interpolation} and \ref{fig:tbs_extrapolation}) for this particular set $({\Delta}E,{\Delta}z,{\Delta}\theta)$. 
As the larger-binned quantity is only an average of a larger bin, it cannot fully account for steeper changes and will fail to reproduce the real value when the gap of missing data is large.

If an entire bin in energy and zenith angle remains with less than two $\alpha$ nodes the hadron-like events are left without correction. 
They are corrected at the end of the analysis chain when the effective template correction $\beta_\mathrm{eff}$ has been calculated for each energy bin (Eq. \ref{eqn:tbs_beta_eff} in the following section). 
In the unlikely event of an energy bin without $\beta_\mathrm{eff}$, the respective excess within the bin in question cannot be calculated and is not considered in the forward folding later on.

\subsection{Effective TBS correction}\label{appendix:eff_tbs}
\begin{figure}[t]
  \resizebox{\hsize}{!}{
    \includegraphics{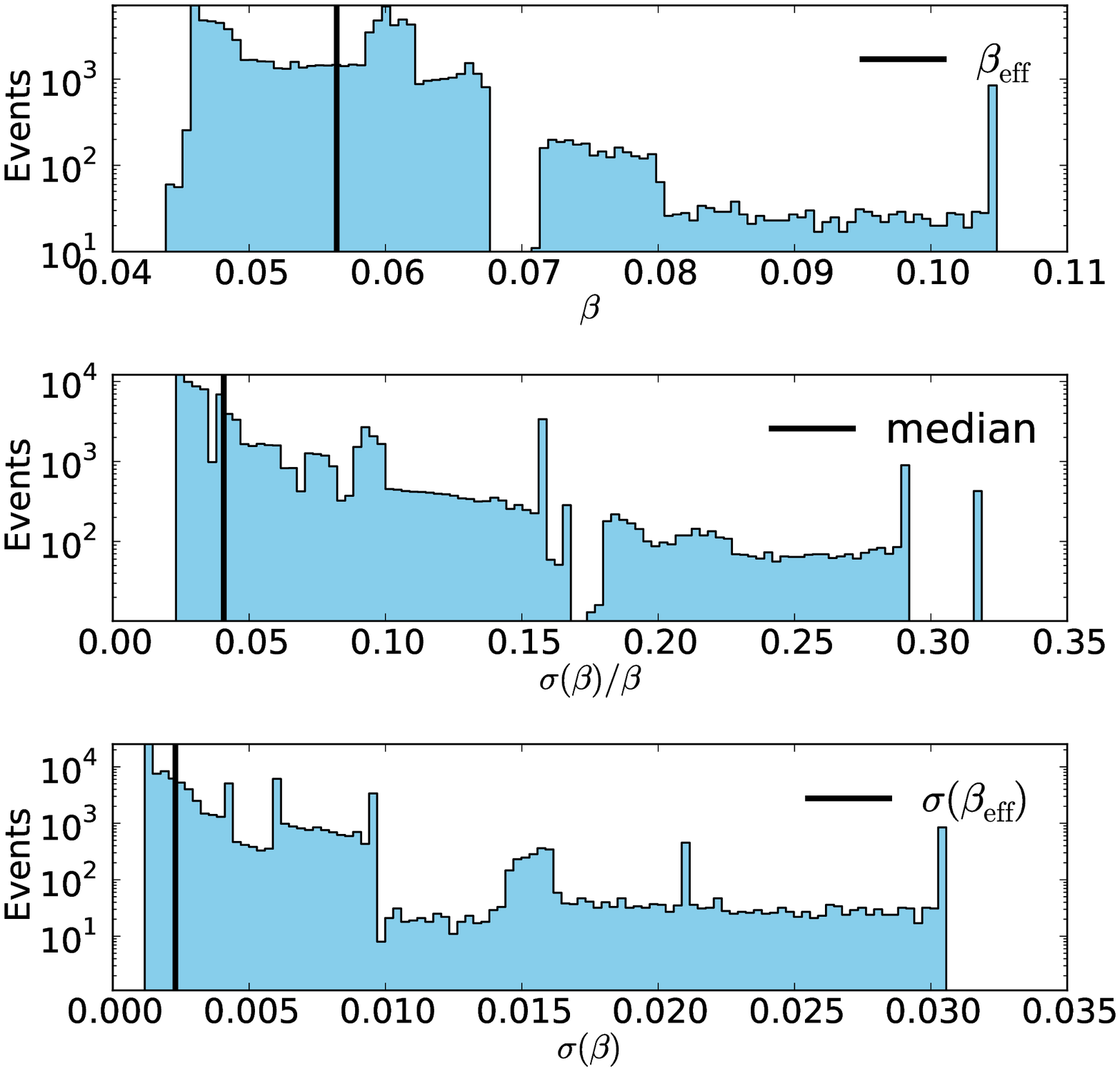}}
  \caption{Sample of $\beta_i$ from \hess\, data on \vx\, \citep{VelaX_paper} for $\Delta{E}=1$ to 1.3\,TeV integrated over the whole zenith-angle and offset range. \emph{Top:} Sample of corrected hadrons $\beta$ after interpolation and extrapolation. The respective value $\beta_\mathrm{eff}$ is marked by the line. \emph{Middle:} Sample of relative errors of the above samples. Here, the black line indicates the median relative error used to calculate $\sigma(\beta_\mathrm{eff})$. \emph{Bottom:} Sample of the errors of the above $\beta$ sample. Marked is the value of $\sigma(\beta_\mathrm{eff})$ which is estimated with the help of the median relative error indicated as a black line in the middle plot. See text for further information.}
  \label{fig:tbs_beta}
\end{figure}
For $m$ hadron-like events from the signal region within an energy interval $\Delta{E}$, every event $i$ with ($z_i$,$\theta_i$) is corrected with the appropriate TBS correction through interpolation or extrapolation (Eqs. \ref{eqn:beta_interpolate2} and \ref{eqn:beta_extrapolate}). 
The effective correction per energy bin $\Delta{E}$ is then
\begin{equation}\label{eqn:tbs_beta_eff}
 \beta_\mathrm{eff}(\Delta{E}) =  \frac{\sum_i^m\beta_i(\Delta{E},z_i,\theta_i)}{N_\mathrm{h}^\mathrm{s}(\Delta{E})}\,.
\end{equation}
Within $\Delta{E}$, the $\beta_i$ are statistically not fully independent as they each depend on at least two variables: $(\alpha_\mathrm{low},\alpha_\mathrm{high})$, $(\alpha_\mathrm{low},\alpha_\mathrm{high},\alpha_\mathrm{shift})$, $(\alpha_\mathrm{next},\alpha_\mathrm{coarse})$, or even $(\alpha_\mathrm{low},\alpha_\mathrm{high},\alpha_\mathrm{shift},\alpha_\mathrm{coarse})$. 
Additionally, the regular nodes $\alpha$, $\alpha_\mathrm{shift}$, and $\alpha_\mathrm{coarse}$ are mostly not independent either. 
An error propagation of Eq. \ref{eqn:tbs_beta_eff} would lead to an underestimate. 

In Fig.\,\ref{fig:tbs_beta}, the distributions of the interpolated and extrapolated quantities $\beta_i$ (\emph{Top} panel), their calculated statistical and relative errors $\sigma_i$ (\emph{Middle} panel) and $\sigma_i/\beta_i$ (\emph{Bottom} panel), respectively, are illustrated for energies between 1\,TeV and 1.3\,TeV. 
The distribution of $\beta_i$ is not continuous but stretched out over a wide range and clustered into populations of different zenith-angle bands. 
The distribution of statistical errors does not follow that of the $\beta_i$ sample. 
The distribution of the relative errors which cover a range of $\sim3-35\,\%$ are used to estimate the error on $\beta_\mathrm{eff}$. 
Conservatively, the error estimate on $\beta_\mathrm{eff}$ per energy bin is defined as
\begin{equation}\label{eqn:tbs_beta_eff_error}
 \sigma(\beta_\mathrm{eff}) =  \beta_\mathrm{eff}\,\mathrm{median}\left(\frac{\sigma(\beta_i)}{\beta_i}\right)\,.
\end{equation}

\section{Exemplary spectrum}\label{crab_spectrum}
\begin{figure}
  \centering
  \resizebox{0.9\hsize}{!}{
    \includegraphics{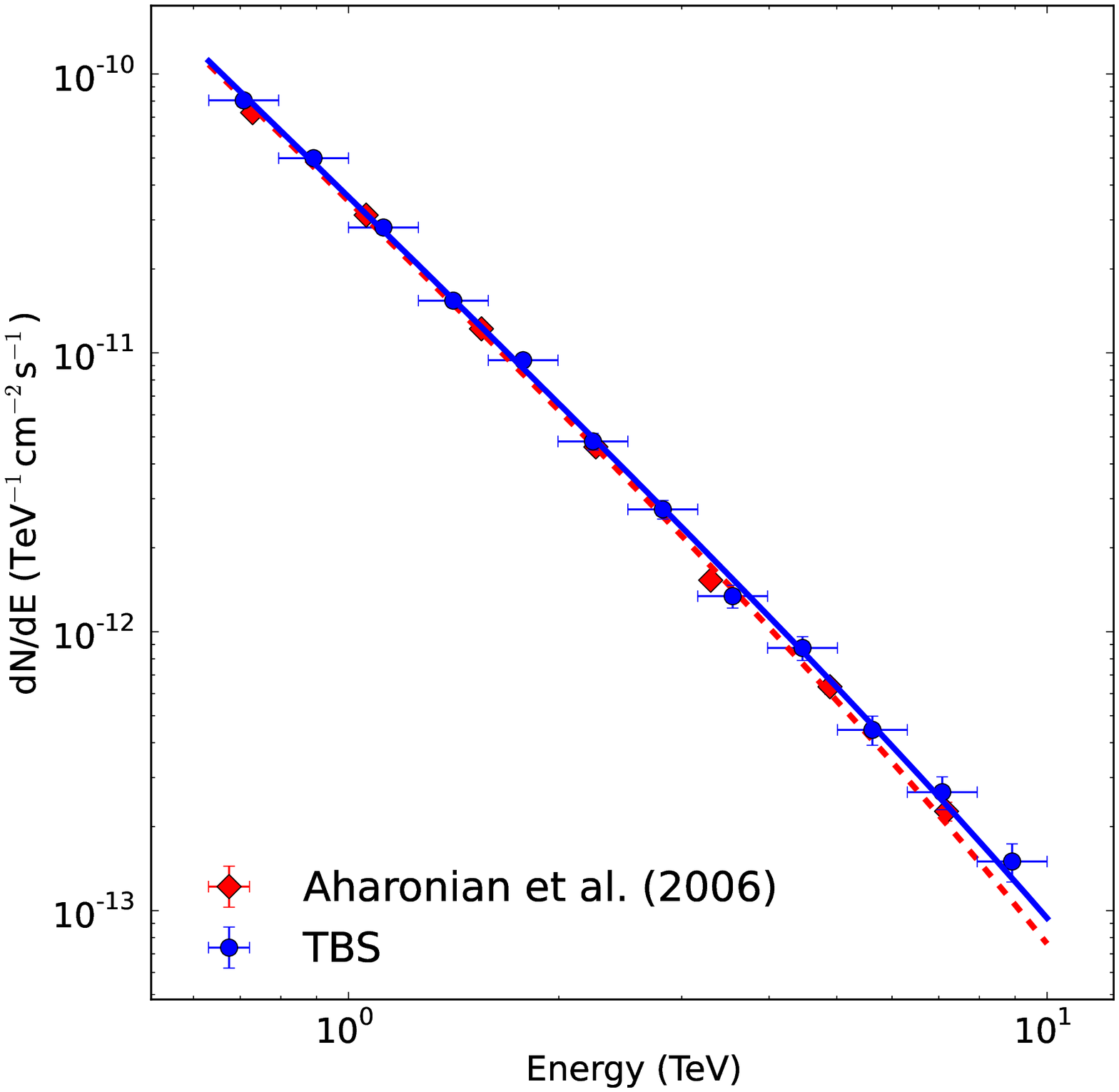}}
  \caption{Differential energy spectra of the \crab\, from \citet[][red-dashed line with red-diamond markers]{Crab_paper} and from this work (TBS; blue-circle markers and blue line indicating the best-fit spectrum). The spectrum above 10\,TeV could not be shown and therefore the spectra were truncated at this energy. See text in Sect.\,\ref{section:discussion} and Appendix \ref{crab_spectrum} for further information.}
  \label{fig:crab}
\end{figure}

In this section, the published \crab\, spectrum \citep[Table\,5 and data set \textit{all} in Table\,6 in][]{Crab_paper} is compared
to the result of TBS in the overlapping energy range. 
The spectrum with the differential flux points published in \citet{Crab_paper} consists of two times more data than the one used here to compare with TBS (see Sect.\,\ref{tab:spectra}).
Hence, the spectrum depicted in Fig.\,\ref{fig:crab} slightly deviates from the one presented earlier.
The energy range above ~10 TeV is subject to ongoing efforts within the \hess\, collaboration related to studies of the source itself and in general studies of the systematics involved at these energies.
Therefore, the last three flux points of the spectrum determined with TBS could not be shown and both spectra were truncated at 10 TeV.
Nevertheless, the spectrum derived with TBS matches that presented in \citet{Crab_paper} very well.

\section{Error contributions}
\begin{figure*}
  \centering
  \hbox{
    \includegraphics[width=0.33\textwidth]{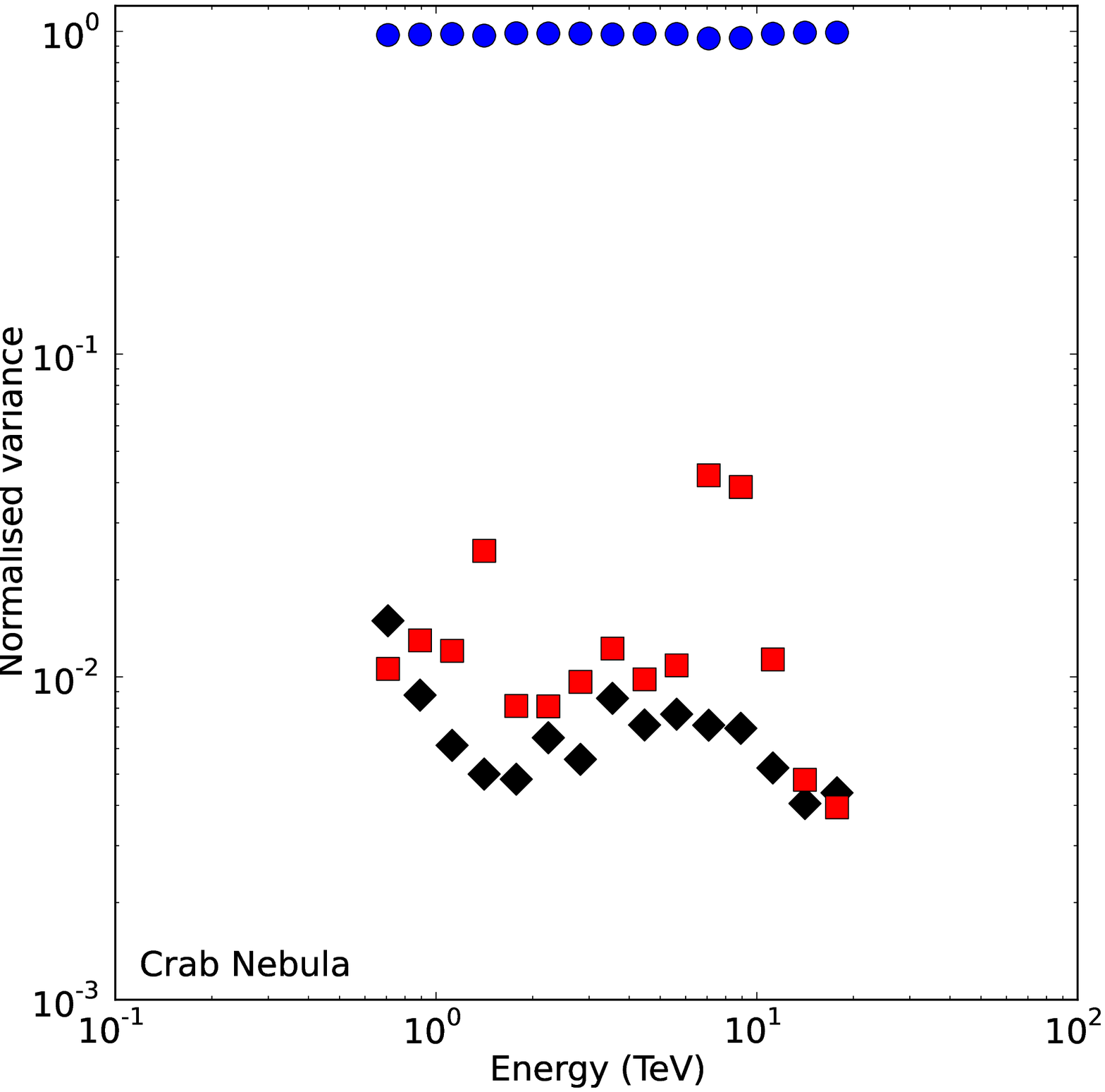}
    \includegraphics[width=0.33\textwidth]{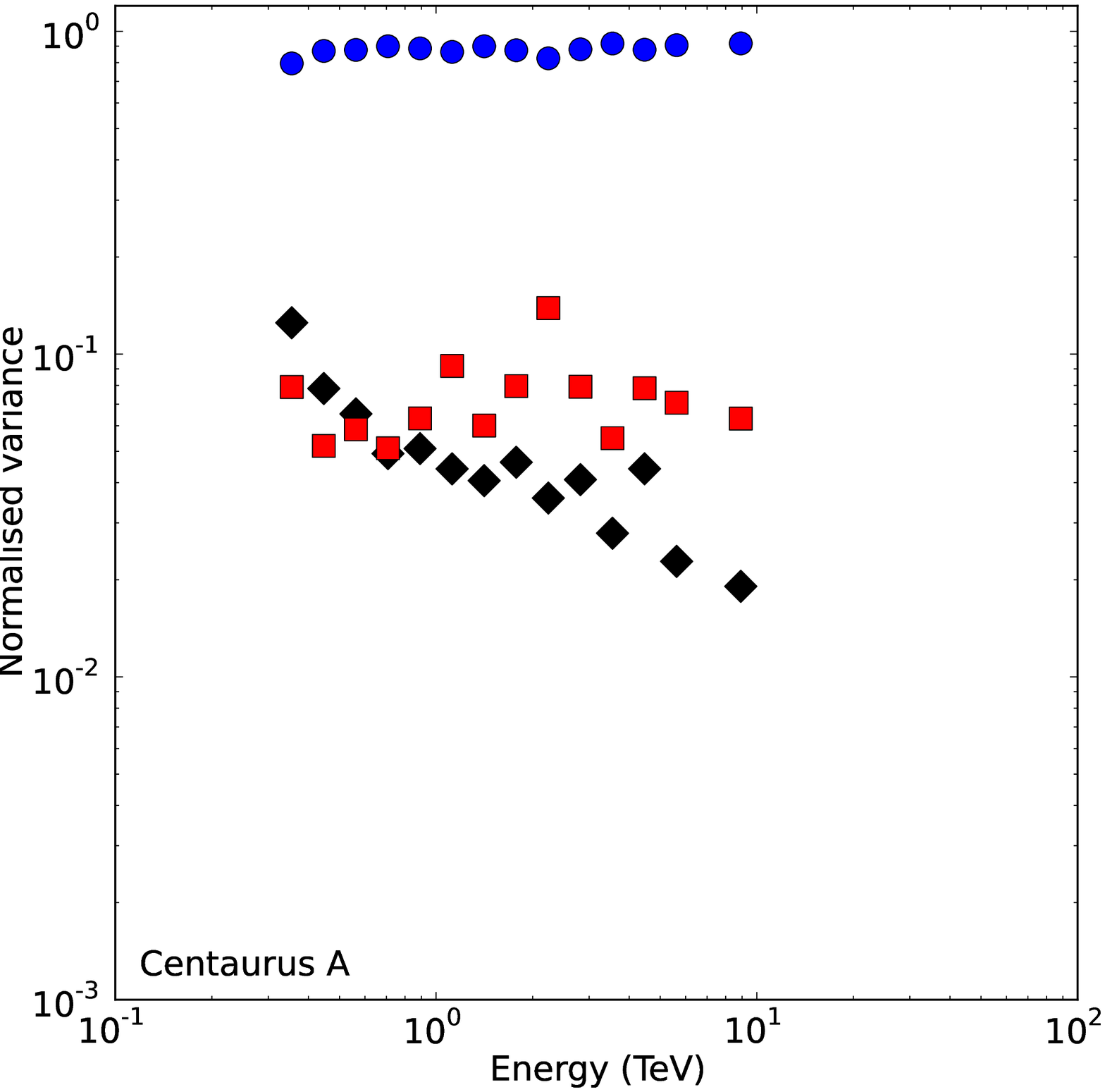}  
    \includegraphics[width=0.33\textwidth]{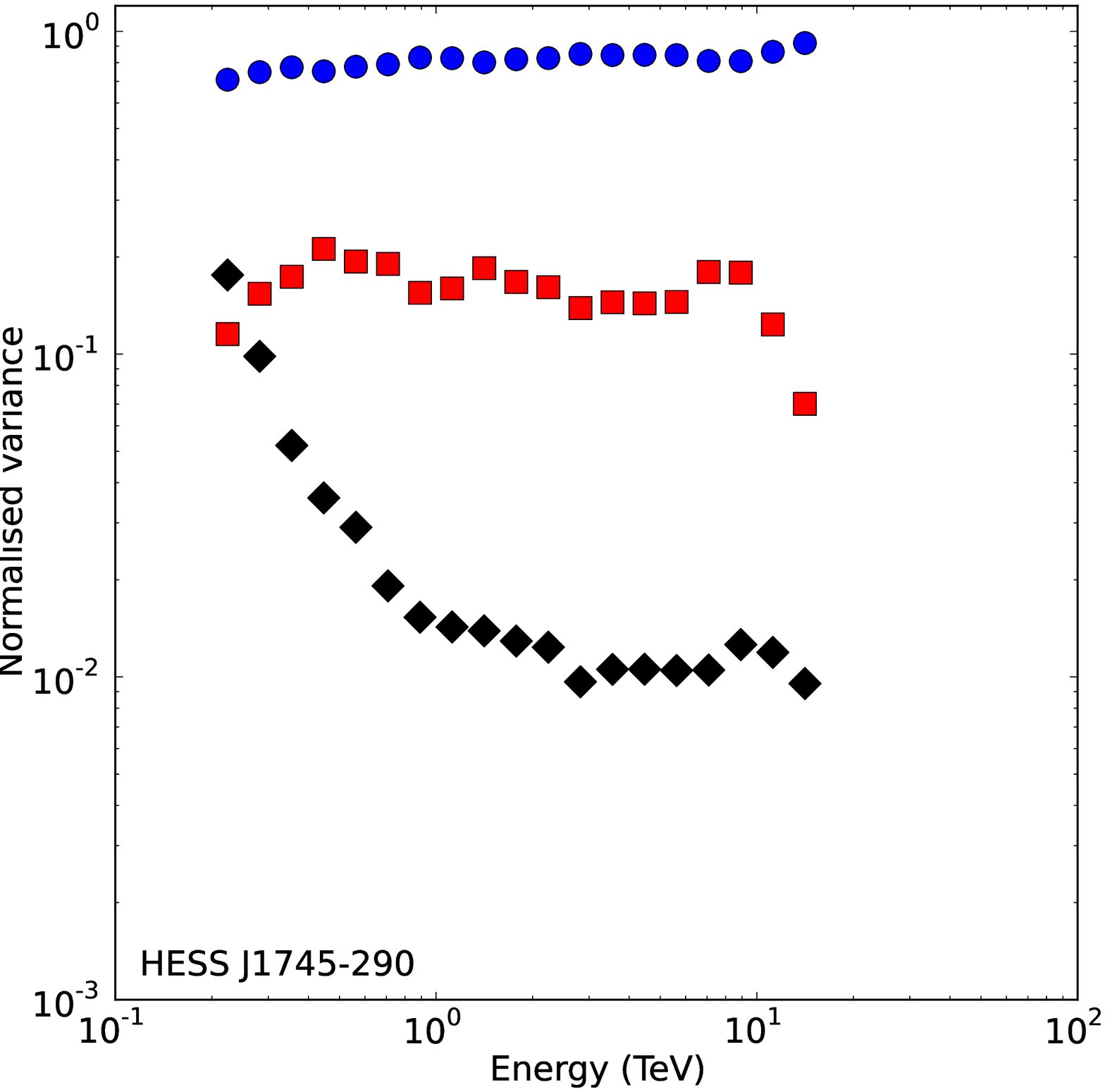}
  }
  \hbox{    
    \includegraphics[width=0.33\textwidth]{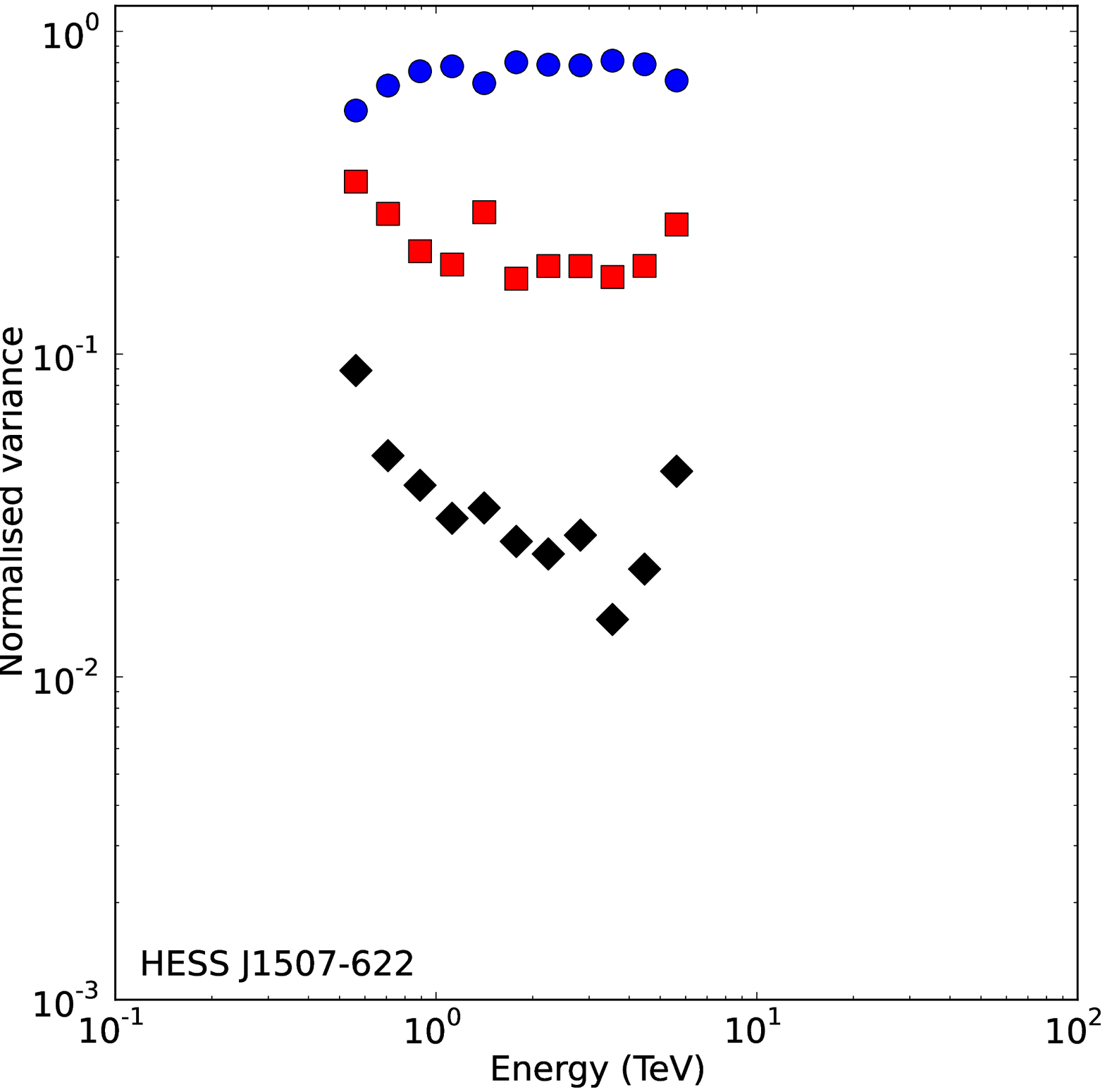}  
    \includegraphics[width=0.33\textwidth]{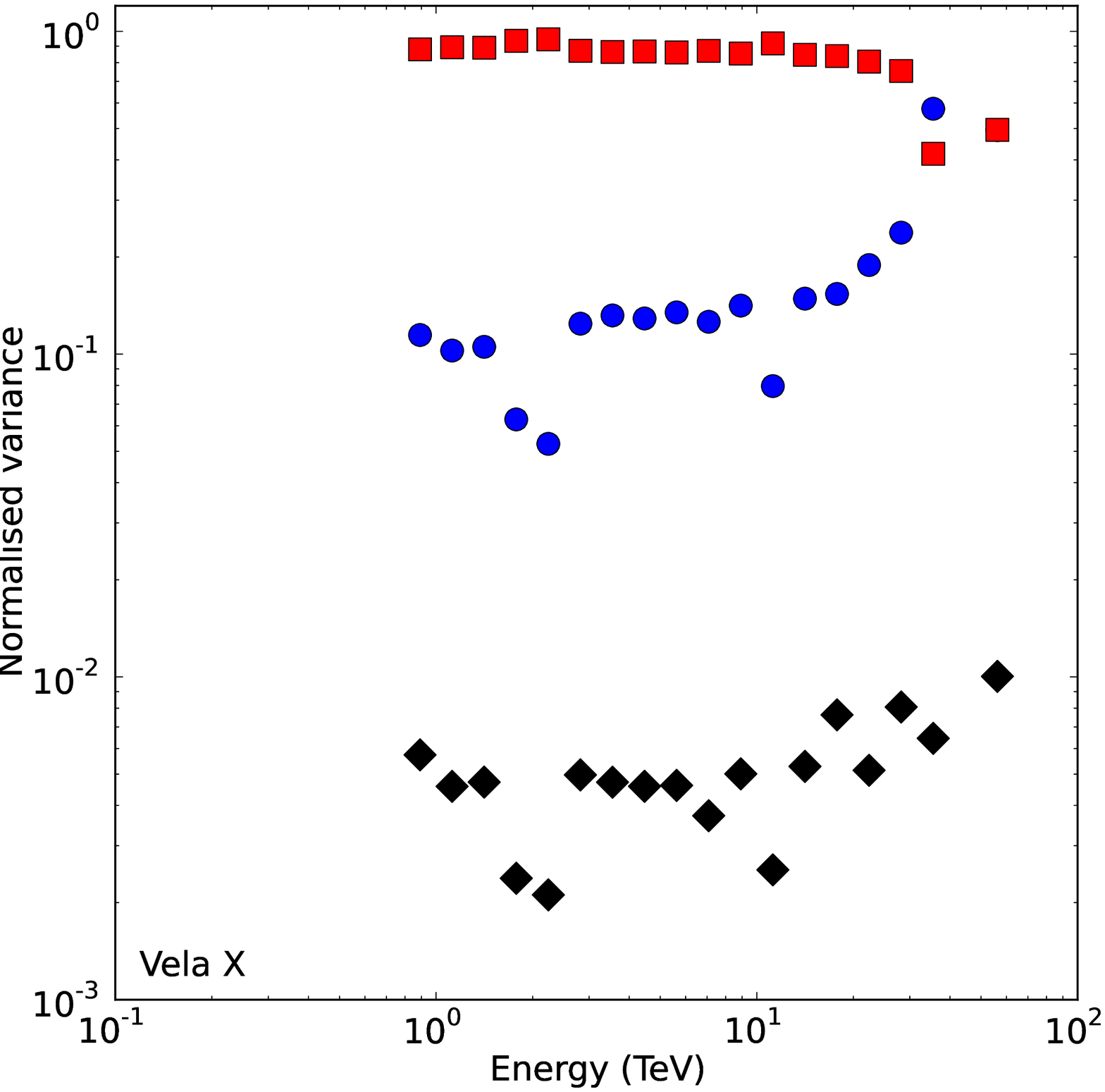}
    \includegraphics[width=0.33\textwidth]{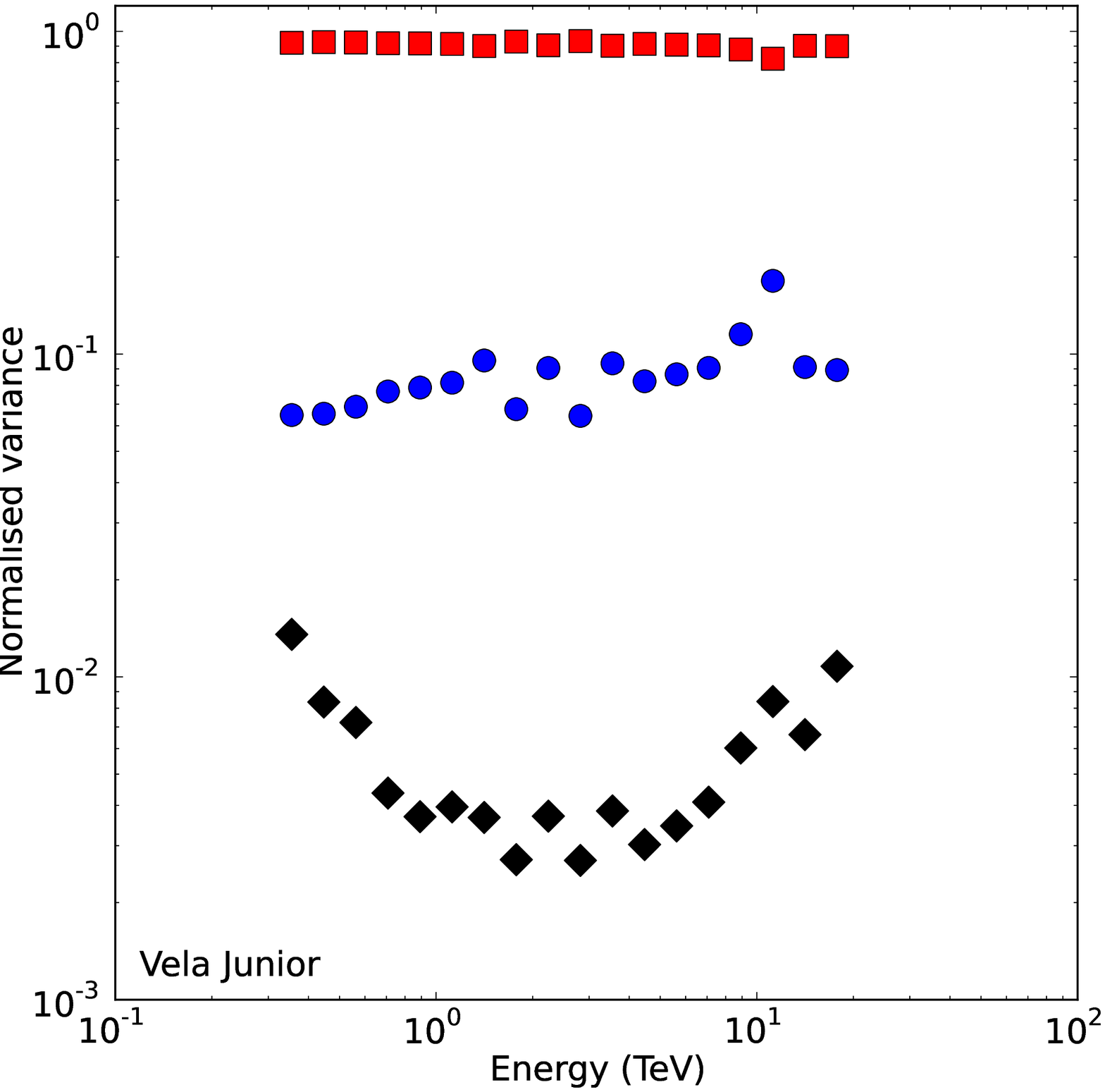}  
  }
  \caption{Normalised variance contribution to Var(e) as defined in Eq. \ref{eqn:var}. For each source Var(g)/Var(e) (blue circles), Var(h)/Var(e) (black diamonds), and Var($\beta_\mathrm{eff}$)/Var(e) (red squares) are shown. See Sect. \ref{section:discussion} for further information and discussion.}
  \label{fig:varplots}
\end{figure*}

In this section, the three different contributions to the overall error on the excess events from TBS are shown for the six analysed sources (Fig.\,\ref{fig:varplots}). 
The discussion is found in Sect.\,\ref{section:discussion}.
\end{appendix}

%    
%__________________________________________________________________

\begin{acknowledgements}
  The authors thank Christian Stegmann and the \hess\, Executive Board for providing access to the data, HAP, and allowing us to show Fig.\,\ref{fig:crab}. MVF acknowledges the financial support from the German Ministry for Education and Research (BMBF, grant no. 05A11GU2). MVF thanks Konrad Bernl\"ohr for providing insight into the H.E.S.S. MCs. MVF thanks Racquel de los Reyes for information on the optical efficiencies for H.E.S.S. MVF thanks Natalie Neumeyer for an interesting discussion on statistics. MVF thanks Ryan Chaves, Phoebe de Wilt, Christoph Deil, Josefa Gonzalez, Helenka Kinnan, Michael Mayer, Franziska Spies, and Christopher van Eldik for their useful comments in different stages of this manuscript. MVF thanks the School of Chemistry and Physics in Adelaide for hosting him for a productive stay in Adelaide. 
  
  This research has made use of NASA's Astrophysics Data System. TBS has been developed using \emph{python} and its modules \emph{matplotlib}, \emph{numpy}, \emph{scipy}, \emph{pyfits}, and \emph{pyminuit}.
\end{acknowledgements}

% for the bibliography, at the end
\bibliographystyle{aa.bst} % style aa.bst
\bibliography{tbs_master.bib} % your references Yourfile.bib
\end{document}